\documentclass[floatfix,aps,prx,twocolumn,groupedaddress,showpacs,english]{revtex4-2}
\usepackage{subcaption}
\usepackage[T1]{fontenc}
\usepackage[dvipsnames]{xcolor}

\usepackage[font=small,labelfont=bf,  justification=justified, format=plain]{caption}
\usepackage[latin9]{inputenc}
\usepackage{amsmath}
\usepackage{graphicx}
\usepackage{esint}

\usepackage{xcolor}
\usepackage{soul}
\renewcommand{\hl}[1]{#1}
\makeatletter
\@ifundefined{showcaptionsetup}{}{%
 \PassOptionsToPackage{caption=false}{subfig}}
\usepackage{subfig}
\usepackage{ragged2e}

\makeatother

\usepackage{babel}
\begin{document}
\title{
Universal framework for uniaxial and biaxial particles with resonance laws and splitting}

\author{Asaf Farhi and Haim Suchowski}
\affiliation{
      School of Physics and Astronomy, Faculty of Exact Sciences, Tel Aviv University, Tel Aviv 69978, Israel}\vspace*{2cm}


\begin{abstract}
Nanophotonics enables precise control over light-matter interactions, though most established design frameworks for subwavelength nanoparticles rely on isotropic materials. Uniaxial and biaxial particles---common in natural and engineered systems---introduce new degrees of freedom coupling geometry and material properties, unlocking multispectral and directional response in previously unexplored spectral regions. We present a universal full-wave framework for eigenmodes and resonances in such nanoparticles. Closed-form solutions reveal axial-permittivity sum rules and anisotropy-induced symmetry breaking, producing resonance splitting and novel radiation patterns. Generalizing to ellipsoids enables geometric tuning of multispectral response, while analytic quality factors elucidate mode localization and loss. Full-wave simulations of h-BN and $\alpha$-MoO3 particles confirm the theory.  This framework unifies the understanding of
 anisotropic nanostructures across optics, magnetism, and thermal transport, opening pathways to a new generation of photonic devices with tunable multispectral response and controlled emission \hl{with direct applications in sensing and imaging}.
\end{abstract}

\maketitle
\paragraph*{Introduction}

Subwavelength structures have attracted significant attention in various fields of physics including photonics, magnetisim, and quantum information,  due to their ability
to provide precise control over the response and produce strong, inherently passive behavior without the need for gain \cite{bergman1979dielectric}.
In photonics, they have been employed to manipulate waves in unprecedented ways, enabling to realize phenomena such as negative refractive index and spasing, as well as advancing applications in sensing, nonlinear emission, and cancer phototherapy \cite{smith2004metamaterials,smith2000negative,bergman2003surface,noginov2009demonstration,kirui2010gold,alu2007epsilon,basov2016polaritons}. While isotropic nanostructures have widespread use \cite{engheta2006metamaterials}, anisotropic nanostructures,  predominantly composed of uniaxial and more recently biaxial materials, are currently at the forefront of photonics research  \cite{tamagnone2018ultra,herzig2024high}.
When particles are composed of anisotropic media, they typically display resonances in unexplored spectral regions, such as in the midinfrared, and exhibit exotic properties that arise from the directional and multispectral nature of the bulk \cite{kim2021extremely}, holding promise for a new generation of anisotropic resonators and biomarkers.
In addition to this class of resonators, a variety of naturally occurring and existing systems are in fact uniaxial particles, ranging from liquid crystal droplets to ferromagnets and ice grains \cite{mishchenko2000light,camenzind2021isotropic,brasselet2009optical,kittel1949physical,breslin2021hyperbolic,korolev2020review}. Moreover, due to the correspondence between the optical properties of isotropic nanoparticles and atoms, better understanding of anisotropic particles may have implications for anisotropic molecules \cite{heugel2010analogy,farhi2024giant,farhi2020three,condon1934absolute}.

When the particle size is much smaller than the wavelength, a quasistatic analysis can be employed. In this regime, 
Laplace's equation plays a dominant role in a variety of fields including electrodynamics, gravity, thermal physics, fluid mechanics, and magnetism \cite{kim2019giant,welp2003magnetic,bairagi2015tuning,lussier1994anisotropy,onofri2010effects,li2003anomalous,hechenblaikner2002direct}. Analytic resonance conditions and eigenstates of isotropic Laplace's equation have been mainly derived for structures with a high degree of symmetry \cite{farhi2014analysis,bergman2014perfect,farhi2017eigenstate,bergman1979dielectric,mayergoyz2013plasmon,koshelev2020dielectric,gladyshev2020symmetry,babicheva2024mie,bohren2008absorption,strattonelectromagnetic,klimov2014nanoplasmonics}. Remarkably, to date, the only structure in physics for which closed-form eigenmode response has been derived is the subwavelength isotropic sphere, which exhibits discrete resonances \cite{stratton2007electromagnetic,bergman1978dielectric,farhi2017eigenstate} (without residual one-dimensional integrals \cite{farhi2014analysis}). Recently, the emergence of anisotropic materials in optics \cite{tamagnone2018ultra,caldwell2019photonics,yu2023hyperbolic,herzig2024high,narimanov2014photonic} has motivated the investigation of the response of anisotropic inclusions, with semianalytical and numerical analyses of spheres and ellipsoids with cartesian anisotropy, eigenmode analysis of a biaxial slab, and studies of spheres with anisotropy in the radial direction \cite{geng2004mie,walker1957magnetostatic,sun2015hamiltonian,alvarez2019analytical,valagiannopoulos2007study,lakhtakia2021theory,lakhtakia2021theory,lakhtakia1991electromagnetic,qiu2007scattering,wallen2015anomalous}. 
In particular, uniaxial and biaxial materials were shown to be very common in semiconductors and crystals, including: SiC (4H), AlN, GaN, quartz, sapphire, calcite, vanadium
oxide (VO$_2$), hexagonal boron nitride (hBN), and $\alpha-$molybdenum trioxide  $(\alpha-$MoO$_3$), making them the most promising among anisotropic particles for a broad range of nanotechnology applications \cite{foteinopoulou2019phonon,herzig2024high,beitner2024localized,breslin2021hyperbolic}. Very recently, an experiment on the modes and resonances of biaxial nanoparticles was performed for the first time by the authors and their colleagues \cite{beitner2024localized}.
However, whereas previous approaches have successfully treated isotropic particles, fundamental challenges arise when analyzing nanostructures with cartesian anisotropy---particularly in predicting higher-order mode behavior---hindering their theoretical understanding. These challenges arise from the mismatch between the particles' geometric and crystal symmetries, the distinct differential equations in each medium, and the additional degrees of freedom introduced by the axial permittivities, which differentiates it from standard eigenvalue problems. This motivates our exploration of a new theoretical framework.
The high-order response of anisotropic particles is important in two common scenarios: when the source is in proximity to the particle \cite{herzig2024high,kurs2007wireless} and when a group of close particles interact e.g., due to a far field excitation, and significantly enhance the field \cite{li2003self,dicke1954coherence}. Hence, addressing the higher-order response is crucial for unlocking this new generation of high-Q tunable, directional, and multispectral resonators and biomarkers that cover unexplored spectral regions, with a variety of novel applications in biomedicine, metamaterials, and photonics \cite{herzig2024high,griffin2018directional,curto2010unidirectional,beitner2024localized}.
\begin{figure}
\includegraphics[width=8cm]{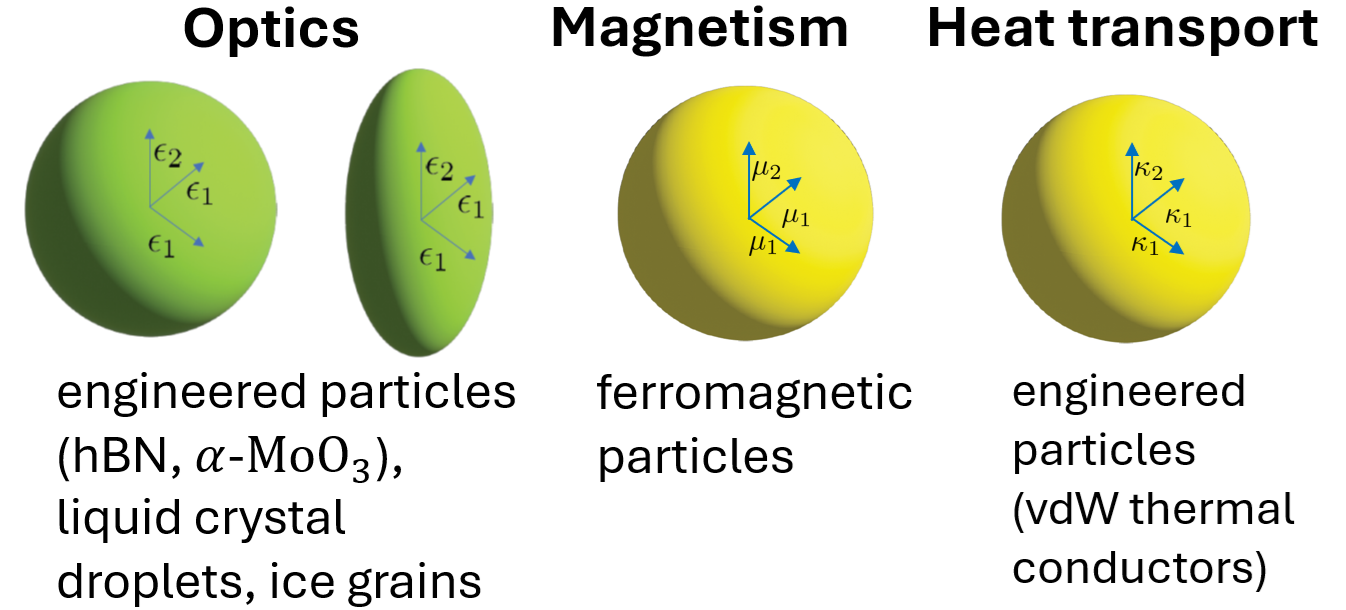}
  \caption{Scope of applicability of the proposed framework }
\end{figure}

\begin{figure*}
\includegraphics[width=16.8cm]{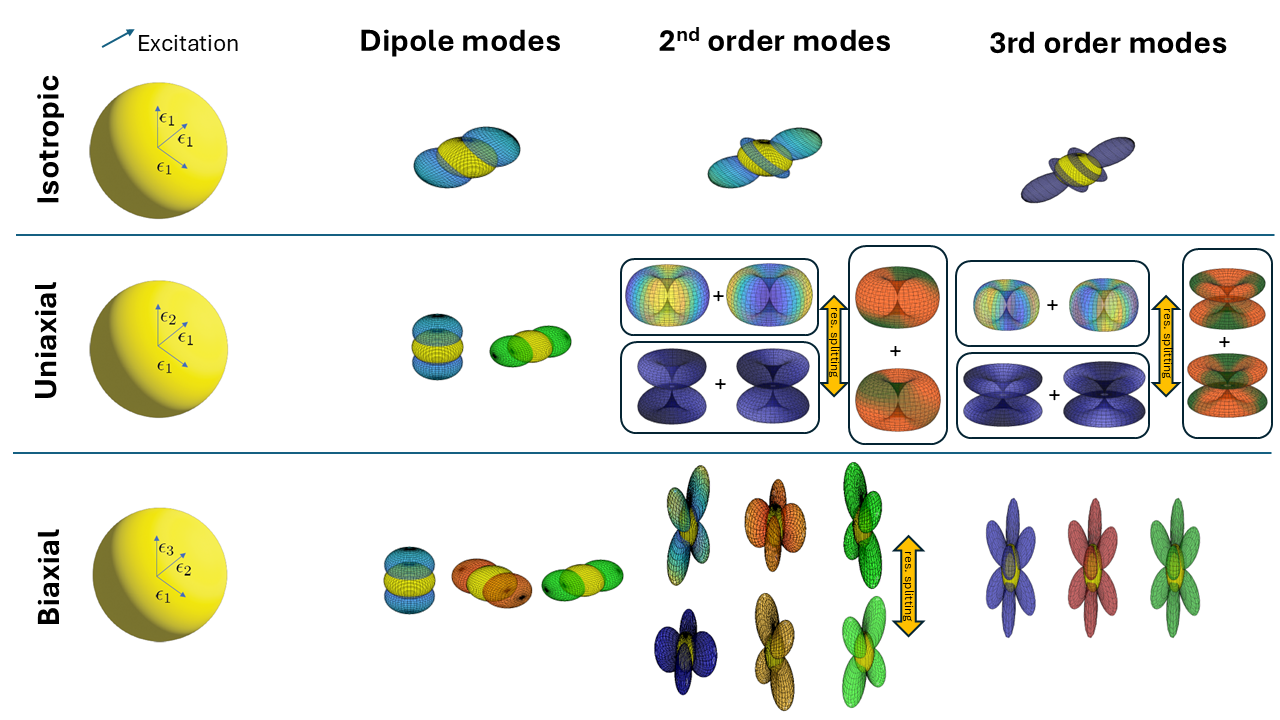}
  \caption{Angular distributions of the quasi-electrostatic potential eigenstates for isotropic, uniaxial, and biaxial spheres. The modes of the isotropic sphere are directed along the field/dipole excitation direction \cite{bergman1978dielectric,farhi2017eigenstate} whereas the modes of the anisotropic spheres are aligned with the crystal axes and multispectral (colors correspond to frequencies). The high-order modes of anisotropic particles exhibit resonance splitting, and degeneracy for uniaxial spheres, leading to  superimposed modes. 
  }
\end{figure*}

Here, we present the second-ever closed-form eigenmode response for a structure, derived for uniaxial and biaxial spheres embedded in an isotropic medium.  
  We identify axial-eigenpermittivity sum
rules, leading to \hl{a new physical phenomenon of anistropic-material induced} resonance splitting, and novel radiation patterns. 
Furthermore, we generalize our results to uniaxial and biaxial ellipsoids, enabling geometric tubability of the response. Moreover, we analytically derive the Q factors for such anisotropic particles.
We apply our theory to uniaxial and biaxial particles composed of hBN and $\alpha-$MoO3 phonon supporting polar crystals, which can be synthesized, and present their unique spectra and eigenmodes  \cite{hillenbrand2002phonon,pyatenko2007synthesis,galanzha2017spaser,okamoto2014fabrication,beitner2024localized}. Finally, we compare our results to full-wave simulations and find excellent agreement. \hl{Critically, our eigenmodes and resonance conditions enable systematic exploration of the ellipsoidal aspect-ratio parameter space for tuning the response to a target frequency in sensing and imaging applications, which is computationally prohibitive using brute-force approaches.} \hl{Our analysis applies to uniaxial and biaxial particles of \emph{any} material and size in optics, magnetism, and heat conduction (Fig. 1), as long as the quasistatic approximation holds.} In addition, it can readily include the temporal dependence of the response by incorporating the time-varying function of the excitation source. Note that whereas previous (semianalytical) works on particles with cartesian anisotropy primarly focused on dielectric materials with all axial permittivities positive $(\epsilon_{1i}>0)$ \cite{lakhtakia2021corrigendum,lakhtakia2021theory}, here we explore the resonant modes that emerge when at least one of the principal components is negative  $(\epsilon_{1i}<0)$. While we consider the localized modes, in practice subwavelength particles radiate and there is a one-to-one mapping between the near and far field modes, which determines the radiation patterns, as we derive in the Supplementary Materail (SM). Even though the radiation of subwavelength particles excited by a far field in free space via high-order modes is negligible, when there is one or more particles or excitation source at a distance$<\lambda$ from the particle, where $\lambda$ is the wavelength, these modes play a  significant role in the radiation.   
\paragraph*{Results} We first derive closed-form eigenmodes and resonance conditions in uniaxial and biaxial spheres. To allow several degrees of freedom of the permittivities in the different axes, we utilize the definition of an eigenstate \cite{bergman1978dielectric,farhi2017eigenstate} as a source-free electric potential in the quasistatic regime applied here to uniaxial and biaxial particles. Our starting point is
 solving Laplace's equation, which reads for anisotropic media \cite{farhi2021coupling}
\[
\nabla\cdot\overleftrightarrow{\epsilon_n}\nabla\psi_{n}=0,
\]
where $\psi_{n}$ is an electric potential eigenstate that exists without a source for the eigenpermittivity tensor  
$\overleftrightarrow{\epsilon_n}=\overleftrightarrow{\epsilon_{1n}}$
inside the inclusion and $\overleftrightarrow{\epsilon_n}=I\epsilon_{2}$
in the host medium ($I$ is the identity matrix). Note that the physical permittivity tensor $\overleftrightarrow{\epsilon_{1}}(\omega)$ generally depends on $\omega$ and the resonances occur when it approximately satisfies the eigenpermittivity relations. Clearly, $\boldsymbol{E}_{n}=\nabla\psi_{n}$ have to satisfy
the boundary conditions at the particle interface. Once anisotropic Laplace's equation and the boundary conditions are satisfied, these functions will be eigenstates or modes of the system by definition. 
In Fig. 2, left column, we present isotropic, uniaxial, and biaxial particles, which are the \emph{most prevalent} types of particles. 

We now derive the eigenmodes and eigenpermittivities of such anisotropic
spheres. We start with the dipole response
and proceed to the high-order mode responses. 
 The anisotropic dipole eigenstates and eigenpermittivities  
can be deduced from the scattering analyses in the literature \cite{bohren2008absorption}
and read for a dipole mode oriented along $z$
\begin{gather}
\tilde{\psi}_{l=1,m=0}=\frac{3}{4\pi}\frac{1}{\sqrt{a}}\left\{ \begin{array}{cc}
\frac{r}{a}\cos\theta & \,r<a\\
\frac{a^{2}}{r^{2}}\cos\theta & r\geq a
\end{array}\right.,\,\,\,\epsilon_{1z}=-2,\nonumber
\end{gather}
where $a$ is the sphere radius. While these eigenstates
and eigenpermittivities appear to be identical to the isotropic ones \cite{bohren2008absorption},
here the modes are oriented along the resonant crystal axis independently
of the incoming field polarization, even though the polarization direction
does affect the excitation magnitude.

We now proceed to the high-order modes of anisotropic spheres. The isotropic eigenstates that conform to the particle geometry (envelope boundary conditions and large-$r$ behavior) \cite{bergman1978dielectric,farhi2017eigenstate} and the eigenstates satisfying cartesian anisotropic electrodynamic equation \cite{alvarez2019analytical} are substantially different---making analytical solutions appear unattainable. Nevertheless, we identify two distinct classes of \emph{joint-property eigenfunctions}. As we consider the geometric requirements to be the more stringent ones, we focus on the isotropic-particle eigenfunctions and impose additional conditions, to show that some of these eigenstates or their superpositions in fact exhibit both properties. Crucially, if these eigenstates satisfy $\partial^2{\psi^m_l}/\partial x_k^2=0$ inside the inclusion, they obey there uniaxial Laplace's equation \hl{since it reduces to isotropic Laplace's equation, namely $\nabla\cdot\overleftrightarrow{\epsilon_n}\nabla\psi^m_{l}=\epsilon_{1n}\nabla^2\psi^m_l= 0.$ As we demonstrate, this approach enables the derivation of the eigenstates and resonance conditions of uniaxial and biaxial subwavelength spheres and ellipsoids.}
Strikingly, functions that satisfy the above condition, are the isotropic-sphere eigenstates $\psi_l^{\pm{(l-1)}},\psi^{\pm{l}}_l,$ \hl{reflecting the removal of degeneracy due to the uniaxial material}, and we obtain their eigenpermittivity relations from the boundary conditions to \emph{arbitrary order}:
\begin{gather}
\tilde{\psi}_{l,u1,2}=\psi_{l}^{\pm\left(l-1\right)}=\left(x\pm iy\right)^{l-1}zf_{l}\left(r\right),\nonumber\\
    \left(l-1\right)\epsilon_{1x}+\epsilon_{1z}=-\epsilon_{2}\left(l+1\right),\,\,\,\epsilon_{1x}=\epsilon_{1y},\nonumber\\
\tilde{\psi}_{l,u3,4}=\psi_{l}^{\pm l}=\left(x\pm iy\right)^{l}f_{l}\left(r\right),\,\,
\epsilon_{1x}=\epsilon_{1y}=-\epsilon_2\frac{l+1}{l},\nonumber\\
f_{l}\left(r\right)=\left\{ \begin{array}{cc}
\frac{1}{\sqrt{la}} (1/a)^l & r<a\\
\frac{1}{\sqrt{la}}  a^{l+1}/r^{2l+1} & r\geq a
\end{array}\right. ,
\end{gather}
where $x,y,z$ correspond to $x_i,x_j,x_k.$ A uniaxial bulk exhibits two resonances with $\epsilon_{1x}=\epsilon_{1y}<0$ and $\epsilon_{1z}<0,$ and one can see that the first condition holds if either $\epsilon_{1x} = \epsilon_{1y} < 0$ or $\epsilon_{1z} < 0$, whereas the second holds only for $\epsilon_{1x} = \epsilon_{1y} < 0.$ \hl{In stark contrast to the isotropic particle case, this results in resonance splitting for $\epsilon_{1x} = \epsilon_{1y} < 0$ since there are two closely-spaced resonances instead of one. Moreover, the first of these resonance conditions typically yields a resonance peak at a lower frequency than the dipole resonance, highlighting a key departure from isotropic particles.}
Also, our modes in both cases are directional as opposed to the isotropic-sphere modes. 
Note that the resonance conditions depend on $\epsilon_2,$ which can be utilized to sense the environment in which the particle is situated.
\begin{figure*}
\includegraphics[width=17cm]{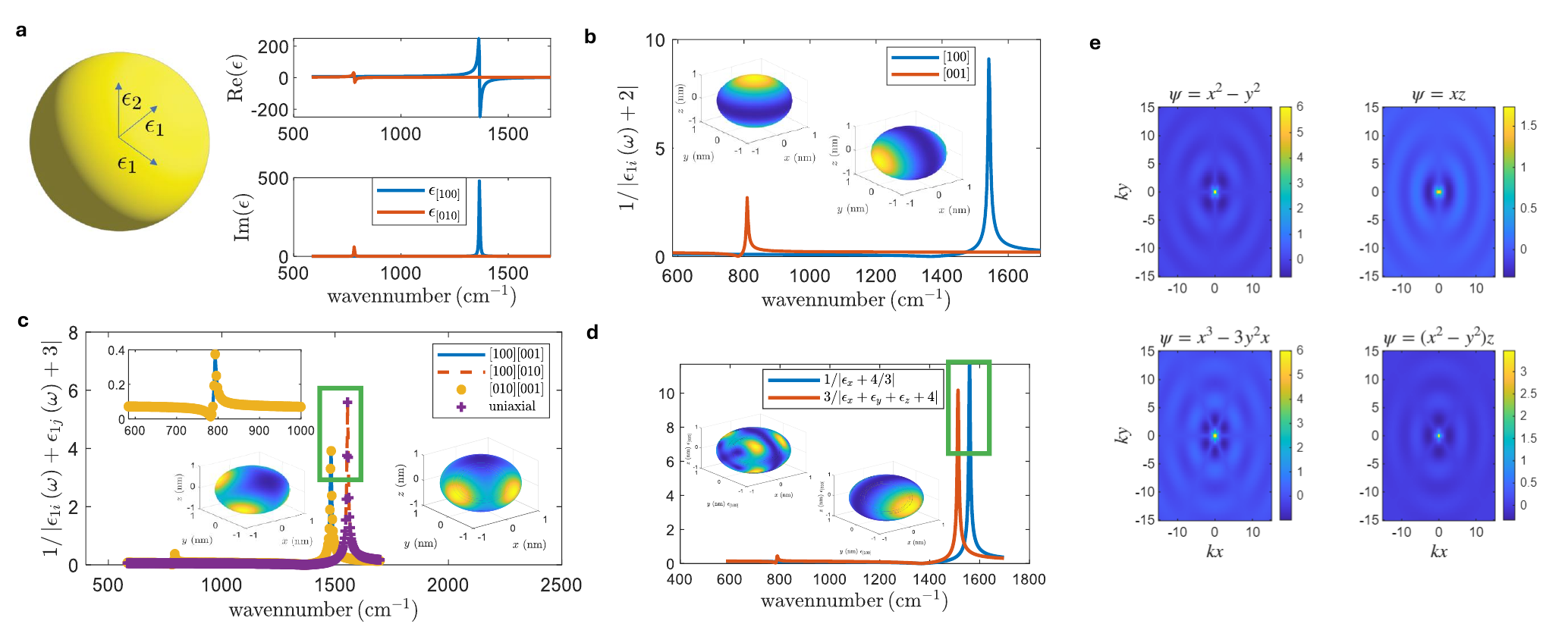}
\caption{Spectrum and eigenmodes of a subwavelength uniaxial sphere. (a) A uniaxial sphere and the permittivity of bulk hBN. (b) The scattering spectrum and $|\mathbf{E}|^2$ of the dipole modes on an hBN sphere surface. Here we see scattering peaks at approximately $820\, \mathrm{(1/cm)}$ (red line) and $1540 \,(\mathrm{1/cm})$ (blue line) associated with  dipole modes in $z$ and $x,$ respectively. The crystal axes [100],[010],[001] are directed along the $x,y,z$ axes, respectively. (c) The scattering spectrum and $|\mathbf{E}|^2$ of the second-order modes, where the $\psi=xz$ and $\psi=yz$ modes and the uniaxial and $\psi=xy$ modes are both doubly degenerate in frequency and  there is a resonance splitting compared to the bulk material. (d) Scattering spectrum and $|\mathbf{E}|^2$ of the third-order  mode, where  $\psi=xyz$ is excited at two $\omega$s. Moreover, there  are geometric-anisotropic $\omega$ shifts compared to the bulk for all modes. Interestingly,  high-order resonances have lower $\omega\mathrm{s}$ compared to the dipole modes, in contrast to isotropic spheres. (e) $|\mathbf{E}|$ of the radiation patterns of four resonances in the $xy$ plane for a dipole source with $\mathbf{r}_0\propto\mathbf{p}\propto(1,1,1)$ resulting in the excitation of the modes $\psi=x^2-y^2,\,\,\psi=xz,\,\,\psi=x^{3}-3y^{2}x,\,\,\,\psi=(x^2-y^2)z.$ }
\end{figure*}

Importantly, there exist superpositions of the isotropic-sphere modes for which $\partial^2 \tilde{\psi}_n / \partial x_k^2 = \partial^2 \tilde{\psi}_n / \partial x_i^2 = 0$ inside the particle. In such cases, the biaxial Laplace equation is \emph{identically satisfied for arbitrary values of} $\epsilon_{1x},,\epsilon_{1y},,\epsilon_{1z}$, reducing the problem solely to imposing boundary conditions. Thus, we derive the following biaxial-sphere modes and calculate the eigenpermittivity relations from the boundary condition: 
\begin{gather}
\tilde{\psi}_{2,b1}=x_ix_jf_l(r),\,\,
\tilde{\psi}_{3,b}=x_ix_jx_kf_l(r),\,
\tilde{\psi}_{4,b1}=x_i^2x_jx_kf_l(r),\nonumber\\
\epsilon_{1i}+\epsilon_{1j}=-3,\,\sum_p \epsilon_{1p}=-4,\,\sum_p \epsilon_{1p}+\epsilon_{1i}=-5. 
\end{gather}
These eigenstates are superpositions of spherical harmonics in all space, as is explained in the SM.
Note that when imposing the boundary condition, this class of functions preserves the functional form under $\partial\psi_{bl}/\partial x_i \hat{r}^i$ inside the particle. 
These sum rules also result in resonance splitting compared to the isotropic-sphere resonances: for a resonance in a given axis in the bulk medium ($\epsilon_{1i}<0$), the sum rules typically result in two resonances occurring at distinct frequencies. 
In the SM we derive eigenstate field expansions for any excitation source and show that the derived uniaxial modes form a complete set and if additional biaxial modes exist they are negligible.
In Fig. 2 we present a scheme of the angular distributions of the electroquasistatic modes of isotropic and anisotropic particles. One can see that anisotropic particles support significantly different modes, which appear in several spectral regions as each axial permittivity experiences a resonance at a different frequency (colors correspond to frequencies).  In addition, their direction is determined only by the crystal orientation. Moreover,  anisotropic particles exhibit rich physics with resonance splitting (illustrated by a slight color change) for both uniaxial and biaxial particles and degeneracy in frequency for uniaxial particles, which leads to  superimposed modes. We generalize this eigenmode derivation to the full-wave regime and obtain the Q factors analytically in the SM.
\begin{figure*}
\includegraphics[width=15.9cm]{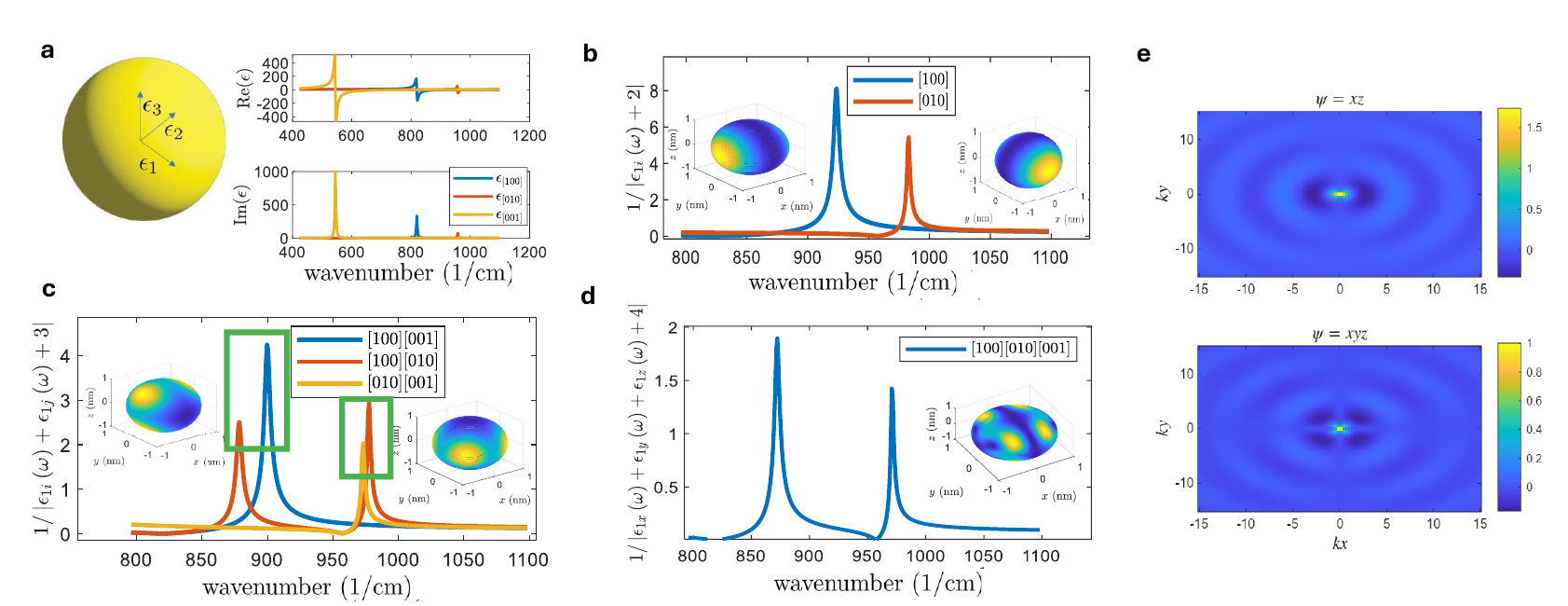}
\caption{Spectrum and eigenmodes of a subwavelength biaxial particle. (a) A biaxial sphere 
 with the bulk permittivities of $\alpha-\mathrm{MoO_3}$. The scattering spectra and $|\mathbf{E}|^2$ of the eigenmodes on a $\alpha-\mathrm{MoO_3}$ sphere surface for the dipole (b), second-order (c), and third-order (d) modes. Note that there are geometric-anisotropic frequency shifts for all modes and resonance splitting for the second-order modes compared to the bulk. (e) $|\mathbf{E}|$ in the radiation patterns of two representative modes of $\psi=xz,\,\,\psi=xyz$ in the $xy$ plane. }
\end{figure*}
 
In Fig. 3 (a) we present a uniaxial sphere and the permittivities of a bulk composed of hBN with two resonances. We then show for a uniaxial sphere composed of hBN the spectrum and $|\mathbf{E}|^2$ on the external surface of the sphere similarly to near-field measurements from the derived resonance conditions and modes, for the dipole (b), second-order (c), and third-order (d) modes. Interestingly, the particle exhibits geometric-anisotropic frequency shifts and the bulk resonance is split for the high-order modes \hl{(green rectangles)}, which also display a degeneracy. \hl{Fig. 3 (e) shows $|\mathbf{E}|$ of the radiation patterns for an oscillating dipole excitation (SM), which are significantly different from those of an isotropic sphere} \cite{farhi2020three}.
In Fig. 4 we present a biaxial sphere and the axial permittivities of $\alpha-\mathrm{MoO_3}$ bulk with three resonances. We then show the calculated scattering spectra and $|\mathbf{E}|^2$ on the sphere surface for the dipole (b), second-order (c), and third-order modes (d). Similarly to the uniaxial particle, the second-order modes exhibit resonance splitting and all the modes display geometric-anisotropic frequency shifts. However, here the uniaxial degeneracy is lifted since there are three different permittivities and the purely-uniaxial modes are not present. Interestingly, in both uniaxial and biaxial spheres, the high-order mode resonances can have lower frequencies compared to the dipole mode, in  contrast to the situation in isotropic spheres.  \hl{Fig. 4 (e) presents $|\mathbf{E}|$ of the radiation patterns of representative biaxial-sphere modes.}

  We now generalize our results to uniaxial and biaxial ellipsoids to enable geometric tunability of the response. Similarly, we suggest that the anisotropic-ellipsoid modes are specific modes of isotropic ellipsoid modes in Niven function basis \cite{guzatov2010plasmon,niven1891vi,whittaker2020course,hobson1931theory}.
  Applying our conditions involving the second derivatives of the potential, we find that the first type of Niven functions fails to satisfy them, whereas the second type does, and corresponds to the modes of the biaxial sphere inside the inclusion.   
We impose continuity of $D_{\perp}$ for the  $l=2$ modes with $\psi_2=x_ix_j$ inside the inclusion by projecting $\boldsymbol{D}$ in the
 direction perpendicular to the interface and equating it with $D_{\perp}$ of the isotropic mode inside the inclusion, since $D_{\perp}$ outside the inclusion is the same in both cases. We thus obtain generalized sum-rule resonance condition:  
\[
a_i^{-2}\epsilon_{1i}+a_j^{-2}\epsilon_{1j}
=\left(a_i^{-2}+a_j^{-2}\right)\epsilon_{1ij}^{\mathrm{iso}}.
\]
where $a_i$ are the ellipsoid semiaxes, and \cite{guzatov2010plasmon} 
\begin{gather}    
\epsilon^\mathrm{iso}_{1ij}=1-\left[\left(\frac{a_{i}a_{j}a_{k}}{2}\right)\left(a_{i}^{2}+a_{j}^{2}\right)I_{ij}\right]^{-1},\nonumber\\
I_{\alpha\beta}=\intop_{0}^{\infty}\frac{du}{\left(u+a_{\alpha}^{2}\right)\left(u+a_{\beta}^{2}\right)R\left(u\right)}.
\end{gather}
For the sphere limit we get $\epsilon_{1i}+\epsilon_{1j}=-3$ as expected. 
We proceed to the third-order mode of an anisotropic ellipsoid. We write the potential and field inside the inclusion $\tilde{\psi}_{3}=xyz.$
Similarly, we equate $D_\perp$ inside the ellipsoid in the isotropic and anisotropic cases and obtain the resonance condition: 
\begin{gather}
a_i^{-2}\epsilon_{1i} + a_j^{-2}\epsilon_{1j} + a_k^{-2}\epsilon_{1k}
= (a_i^{-2}+a_j^{-2}+a_k^{-2})\,\epsilon_{3}^{\mathrm{iso}},\\
\epsilon_{3}^{\mathrm{iso}}=1-\left[\frac{\left(a_{1}a_{2}a_{3}\right)^{3}}{2}\left(\sum_{\alpha=1}^{3}a_{\alpha}^{-2}I_{123}\right)\right]^{-1},I_{123}=\intop_{0}^{\infty}\frac{du}{R^{3}\left(u\right)}.\nonumber
\end{gather}
We take the anisotropic-sphere limit of the  eigenpermittivity relation and obtain agreement
$ \epsilon_{1x}+\epsilon_{1y}+\epsilon_{1z}=3\epsilon_{3\,\mathrm{iso}}=-4. $ In this way one can readily find the fourth-order biaxial ellipsoid mode and resonance condition.
Finally, we verified our theory for uniaxial and biaxial spheres and a biaxial ellipsoid in extensive hyperspectral full-wave simulations with excellent agreement for the modes,  spectra, Q factors, and radiation patterns in the SM. In our recent work in Ref. \cite{beitner2024localized} we also verified the second-order mode response experimentally for two biaxial ellipsoids with hyperspectral mid-IR near-field imaging. 
\paragraph*{Discussion}
We derived the 
modes and eigenpermittivity relations of uniaxial and biaxial 
 nanospheres in the quasistatic regime as well as their radiation patterns in closed form. We showed that while isotropic spheres
have discrete eigenpermittivities,  anisotropic spheres exhibit eigenpermittivity sum rules that couple 
the axial permittivities, uncovering to a \hl{novel physical phenomenon of anisotropic-material induced} resonance splitting. We also identified eigenfrequency degeneracy for uniaxial spheres, resulting in superimposed modes. Interestingly, as opposed to isotropic-material particles, high-order resonance frequencies in anisotropic particles are often lower than the dipole resonance.
We generalized our results to uniaxial and biaxial ellipsoids, with eigenpermittivity sum rules that account for both the anisotropic material properties and the ellipsoid's geometry, allowing tunability of the response. Finally, we derived the Q factors analytically and performed extensive full-wave simulations that agreed very well with \hl{ our derived modes, resonance conditions, resonance widths, and excitation-property dependency--- the key components of our eigenstate expansion}. 

\hl{
Our analysis provides a unified foundation for controlling light-matter interactions in anisotropic particles, with direct implications for thermal emission, spontaneous emission enhancement, and nanoscale energy transfer} \cite{carminati1999near,biehs2021near,purcell1995spontaneous,dung2002intermolecular}. \hl{ Moreover, our modes and permittivity relations could be utilized to construct hybridized eigenstates and resonances conditions for an array of such nanoparticles} \cite{bergman1979dielectric,li2003self}. \hl{While geometric anisotropy alone already induces directional response in isotropic ellipsoids, it typically leads to uneven modal strengths and spectrally narrow features, particularly suppressing modes associated with the long axes. In contrast, material anisotropy leads to a more balanced distribution of modal strength across modes, enabling multiple pronounced resonances with broad spectral support arising from the directional and multispectral nature of the bulk. Moreover, anisotropic materials exhibit resonances in spectral regions distinct from those of isotropic materials, further distinguishing the properties of the corresponding particles. }

\hl{
 Notably, the resonance rules for anisotropic ellipsoids enable systematic tuning of spectral responses into previously unexplored mid-infrared regions, which is crucial for sensing and imaging applications requiring precise matching to molecular vibrational bands. Importantly, anisotropic nanoparticles allow for capabilities such as spatially resolved response and enhancement within the sample that are not achievable with other geometries. Specifically, the ability to compute the full resonance spectrum of a \emph{given particle} immediately from the analytic expressions---rather than relying on day-long full-wave simulations---opens the door to rapid exploration and inverse design across the entire ellipsoidal aspect-ratio parameter space. Strikingly, our framework completes an exploration of this parameter space in seconds that would otherwise require several years of full-wave simulation or weeks using standard numerical static solvers. Moreover, when considering different illumination directions or source locations, finite-element simulations must be rerun for each configuration. In contrast, our approach only requires recalculating the mode coefficients, resulting in an additional multiplicative improvement in computational cost. Experimental implementation of sensing can be performed via wide-field or focused illumination leading to heating of bound targets} \cite{kirui2010gold,galanzha2017spaser}, \hl{acoustic emission} \cite{galanzha2017spaser}, \hl{or excitation of high-order particle-target multipole channels, which arise only in particle-target proximity} \cite{farhi2024giant} \hl{The resulting multispectral and environment-sensitive behavior enables highly specific spectral signatures and simulataneous multi-protein detection, positioning anisotropic nanoparticles as promising candidates for biomarkers and for probing complex biological environments, with potential implications for early cancer detection} \cite{lee2024super}. Similarly, one could rapidly calculate the Q factors for the entire hyperspectral aspect-ratio parameter space using our derived resonance conditions, an endeavor that would otherwise be computationally prohibitive.
 
\hl{Furthermore, because thermal radiation predominantly occupies the mid-IR spectral range, anisotropic nanoparticles may enable thermal sensors that map incoming thermal radiation onto multiple spectral and spatial channels. The tailored radiation patterns enabled by anisotropy suggest opportunities in directional emission control and nanoscale light routing, as well as potential applications in detection schemes including for dark matter} \cite{griffin2018directional,kadribasic2018directional,boyd2023directional}. 
\hl{
These resonant modes also influence quantum emission processes while their strong field localization and spectral multiplicity, governed by the underlying sum rules, provide a fertile platform for nonlinear optics and optical information processing. Experimentally, anisotropic spheres and ellipsoids have been synthesizied using laser ablation in acetone and superfluid helium with shape control via imaging filtering and adjusting temperature or surface tension and pressure of the liquid helium, respectively} \cite{okamoto2014fabrication,beitner2024localized}. 
\hl{More broadly, the generality of our framework extends beyond photonics, offering new avenues for engineering resonant phenomena in magnetism and thermal transport } \cite{kurs2007wireless}.

\paragraph*{Acknowledgement}
E. E. Narimanov, A. Niv, Y. Mazor, J. Jeffet, Y. Reches, and I. Kalyan are acknowledged for the insightful comments.
We acknowledge the support from the ISF grant 2312/20 and a grant from The Ministry of Aliyah and Integration.

\bibliographystyle{unsrt}

\begin{thebibliography}{10}

\bibitem{bergman1979dielectric}
David~J Bergman.
\newblock Dielectric constant of a two-component granular composite: A
  practical scheme for calculating the pole spectrum.
\newblock {\em Physical Review B}, 19(4):2359, 1979.

\bibitem{smith2004metamaterials}
David~R Smith, John~B Pendry, and Mike~CK Wiltshire.
\newblock Metamaterials and negative refractive index.
\newblock {\em science}, 305(5685):788--792, 2004.

\bibitem{smith2000negative}
David~R Smith and Norman Kroll.
\newblock Negative refractive index in left-handed materials.
\newblock {\em Physical review letters}, 85(14):2933, 2000.

\bibitem{bergman2003surface}
David~J Bergman and Mark~I Stockman.
\newblock Surface plasmon amplification by stimulated emission of radiation:
  Quantum generation of coherent surface plasmons in nanosystems.
\newblock {\em Physical review letters}, 90(2):027402, 2003.

\bibitem{noginov2009demonstration}
MA~Noginov, G~Zhu, AM~Belgrave, Reuben Bakker, VM~Shalaev, EE~Narimanov,
  S~Stout, E~Herz, T~Suteewong, and U~Wiesner.
\newblock Demonstration of a spaser-based nanolaser.
\newblock {\em Nature}, 460(7259):1110--1112, 2009.

\bibitem{kirui2010gold}
Dickson~K Kirui, Diego~A Rey, and Carl~A Batt.
\newblock Gold hybrid nanoparticles for targeted phototherapy and cancer
  imaging.
\newblock {\em Nanotechnology}, 21(10):105105, 2010.

\bibitem{alu2007epsilon}
Andrea Alu, Mario~G Silveirinha, Alessandro Salandrino, and Nader Engheta.
\newblock Epsilon-near-zero metamaterials and electromagnetic sources:
  Tailoring the radiation phase pattern.
\newblock {\em Physical Review B}, 75(15):155410, 2007.

\bibitem{basov2016polaritons}
DN~Basov, MM~Fogler, and FJ~Garcia~de Abajo.
\newblock Polaritons in van der waals materials.
\newblock {\em Science}, 354(6309):aag1992, 2016.

\bibitem{engheta2006metamaterials}
Nader Engheta and Richard~W Ziolkowski.
\newblock {\em Metamaterials: physics and engineering explorations}.
\newblock John Wiley \& Sons, 2006.

\bibitem{tamagnone2018ultra}
Michele Tamagnone, Antonio Ambrosio, Kundan Chaudhary, Luis~A Jauregui, Philip
  Kim, William~L Wilson, and Federico Capasso.
\newblock Ultra-confined mid-infrared resonant phonon polaritons in van der
  waals nanostructures.
\newblock {\em Science advances}, 4(6):7189, 2018.

\bibitem{herzig2024high}
Hanan Herzig~Sheinfux, Lorenzo Orsini, Minwoo Jung, Iacopo Torre, Matteo
  Ceccanti, Simone Marconi, Rinu Maniyara, David Barcons~Ruiz, Alexander
  H{\"o}tger, Ricardo Bertini, et~al.
\newblock High-quality nanocavities through multimodal confinement of
  hyperbolic polaritons in hexagonal boron nitride.
\newblock {\em Nature Materials}, 23(4):499--505, 2024.

\bibitem{kim2021extremely}
Shi~En Kim, Fauzia Mujid, Akash Rai, Fredrik Eriksson, Joonki Suh, Preeti
  Poddar, Ariana Ray, Chibeom Park, Erik Fransson, Yu~Zhong, et~al.
\newblock Extremely anisotropic van der waals thermal conductors.
\newblock {\em Nature}, 597(7878):660--665, 2021.

\bibitem{mishchenko2000light}
Michael~I Mishchenko, Joop~W Hovenier, and Larry~D Travis.
\newblock Light scattering by nonspherical particles: theory, measurements, and
  applications.
\newblock {\em Measurement Science and Technology}, 11(12):1827--1827, 2000.

\bibitem{camenzind2021isotropic}
Leon~C Camenzind, Simon Svab, Peter Stano, Liuqi Yu, Jeramy~D Zimmerman,
  Arthur~C Gossard, Daniel Loss, and Dominik~M Zumb{\"u}hl.
\newblock Isotropic and anisotropic g-factor corrections in gaas quantum dots.
\newblock {\em Physical Review Letters}, 127(5):057701, 2021.

\bibitem{brasselet2009optical}
Etienne Brasselet, Naoki Murazawa, Hiroaki Misawa, and Saulius Juodkazis.
\newblock Optical vortices from liquid crystal droplets.
\newblock {\em Physical review letters}, 103(10):103903, 2009.

\bibitem{kittel1949physical}
Charles Kittel.
\newblock Physical theory of ferromagnetic domains.
\newblock {\em Reviews of modern Physics}, 21(4):541, 1949.

\bibitem{breslin2021hyperbolic}
Vanessa~M Breslin, Daniel~C Ratchford, Alexander~J Giles, Adam~D Dunkelberger,
  and Jeffrey~C Owrutsky.
\newblock Hyperbolic phonon polariton resonances in calcite nanopillars.
\newblock {\em Optics Express}, 29(8):11760--11772, 2021.

\bibitem{korolev2020review}
Alexei Korolev and Thomas Leisner.
\newblock Review of experimental studies of secondary ice production.
\newblock {\em Atmospheric Chemistry and Physics}, 20(20):11767--11797, 2020.

\bibitem{heugel2010analogy}
Simon Heugel, Alessandro~S Villar, Markus Sondermann, Ulf Peschel, and Gerd
  Leuchs.
\newblock On the analogy between a single atom and an optical resonator.
\newblock {\em Laser Physics}, 20:100--106, 2010.

\bibitem{farhi2024giant}
Asaf Farhi.
\newblock Giant enhancement of high-order rotational and rovibrational
  transitions in near-field spectroscopy in proximity to nanostructures.
\newblock {\em Physical Review Applied}, 21(3):034047, 2024.

\bibitem{farhi2020three}
Asaf Farhi.
\newblock Three-dimensional-subwavelength field localization, time reversal of
  sources, and infinite, asymptotic degeneracy in spherical structures.
\newblock {\em Physical Review A}, 101(6):063818, 2020.

\bibitem{condon1934absolute}
EU~Condon.
\newblock The absolute intensity of the nebular lines.
\newblock {\em Astrophysical Journal, vol. 79, p. 217}, 79:217, 1934.

\bibitem{kim2019giant}
Dong-Hwan Kim, Kyoo Kim, Kyung-Tae Ko, JunHo Seo, Jun~Sung Kim, Tae-Hwan Jang,
  Younghak Kim, Jae-Young Kim, Sang-Wook Cheong, and Jae-Hoon Park.
\newblock Giant magnetic anisotropy induced by ligand l s coupling in layered
  cr compounds.
\newblock {\em Physical review letters}, 122(20):207201, 2019.

\bibitem{welp2003magnetic}
U~Welp, VK~Vlasko-Vlasov, X~Liu, JK~Furdyna, and T~Wojtowicz.
\newblock Magnetic domain structure and magnetic anisotropy in
  $\mathrm{Ga_{1-x}Mn_x As}$.
\newblock {\em Physical review letters}, 90(16):167206, 2003.

\bibitem{bairagi2015tuning}
Kaushik Bairagi, Amandine Bellec, Vincent Repain, C~Chacon, Y~Girard, Yves
  Garreau, J~Lagoute, S~Rousset, R~Breitwieser, Yu-Cheng Hu, et~al.
\newblock Tuning the magnetic anisotropy at a molecule-metal interface.
\newblock {\em Physical review letters}, 114(24):247203, 2015.

\bibitem{lussier1994anisotropy}
Benoit Lussier, Brett Ellman, and Louis Taillefer.
\newblock Anisotropy of heat conduction in the heavy fermion superconductor u
  pt 3.
\newblock {\em Physical review letters}, 73(24):3294, 1994.

\bibitem{onofri2010effects}
M~Onofri, F~Malara, and Pierluigi Veltri.
\newblock Effects of anisotropic thermal conductivity in magnetohydrodynamics
  simulations of a reversed-field pinch.
\newblock {\em Physical review letters}, 105(21):215006, 2010.

\bibitem{li2003anomalous}
Baowen Li and Jiao Wang.
\newblock Anomalous heat conduction and anomalous diffusion in one-dimensional
  systems.
\newblock {\em Physical review letters}, 91(4):044301, 2003.

\bibitem{hechenblaikner2002direct}
Gerald Hechenblaikner, Eleanor Hodby, Stephen~A Hopkins, Onofrio~M Marago, and
  Christopher~J Foot.
\newblock Direct observation of irrotational flow and evidence of superfluidity
  in a rotating bose-einstein condensate.
\newblock {\em Physical review letters}, 88(7):070406, 2002.

\bibitem{farhi2014analysis}
Asaf Farhi and David~J Bergman.
\newblock Analysis of a veselago lens in the quasistatic regime.
\newblock {\em Physical Review A}, 90(1):013806, 2014.

\bibitem{bergman2014perfect}
David~J Bergman.
\newblock Perfect imaging of a point charge in the quasistatic regime.
\newblock {\em Physical Review A}, 89(1):015801, 2014.

\bibitem{farhi2017eigenstate}
Asaf Farhi and David~J Bergman.
\newblock Eigenstate expansion of the quasistatic electric field of a point
  charge in a spherical inclusion structure.
\newblock {\em Physical Review A}, 96(4):043806, 2017.

\bibitem{mayergoyz2013plasmon}
Isaak~D Mayergoyz.
\newblock {\em Plasmon resonances in nanoparticles}, volume~6.
\newblock World Scientific, 2013.

\bibitem{koshelev2020dielectric}
Kirill Koshelev and Yuri Kivshar.
\newblock Dielectric resonant metaphotonics.
\newblock {\em Acs Photonics}, 8(1):102--112, 2020.

\bibitem{gladyshev2020symmetry}
Sergey Gladyshev, Kristina Frizyuk, and Andrey Bogdanov.
\newblock Symmetry analysis and multipole classification of eigenmodes in
  electromagnetic resonators for engineering their optical properties.
\newblock {\em Physical Review B}, 102(7):075103, 2020.

\bibitem{babicheva2024mie}
Viktoriia~E Babicheva and Andrey~B Evlyukhin.
\newblock Mie-resonant metaphotonics.
\newblock {\em Advances in Optics and Photonics}, 16(3):539--658, 2024.

\bibitem{bohren2008absorption}
Craig~F Bohren and Donald~R Huffman.
\newblock {\em Absorption and scattering of light by small particles}.
\newblock John Wiley \& Sons, 2004.

\bibitem{strattonelectromagnetic}
JA~Stratton.
\newblock {\em Electromagnetic theory (1941), New York}.
\newblock McGraw-Hill.

\bibitem{klimov2014nanoplasmonics}
Vasily Klimov.
\newblock {\em Nanoplasmonics}.
\newblock CRC press, 2014.

\bibitem{stratton2007electromagnetic}
Julius~Adams Stratton.
\newblock {\em Electromagnetic theory}, volume~33.
\newblock John Wiley \& Sons, 2007.

\bibitem{bergman1978dielectric}
David~J Bergman.
\newblock The dielectric constant of a composite material-a problem in
  classical physics.
\newblock {\em Physics Reports}, 43(9):377--407, 1978.

\bibitem{caldwell2019photonics}
Joshua~D Caldwell, Igor Aharonovich, Guillaume Cassabois, James~H Edgar,
  Bernard Gil, and DN~Basov.
\newblock Photonics with hexagonal boron nitride.
\newblock {\em Nature Reviews Materials}, 4(8):552--567, 2019.

\bibitem{yu2023hyperbolic}
Shang-Jie Yu, Helen Yao, Guangwei Hu, Yue Jiang, Xiaolin Zheng, Shanhui Fan,
  Tony~F Heinz, and Jonathan~A Fan.
\newblock Hyperbolic polaritonic rulers based on van der waals $\alpha$-moo3
  waveguides and resonators.
\newblock {\em ACS nano}, 17(22):23057--23064, 2023.

\bibitem{narimanov2014photonic}
Evgenii~E Narimanov.
\newblock Photonic hypercrystals.
\newblock {\em Physical Review X}, 4(4):041014, 2014.

\bibitem{geng2004mie}
You-Lin Geng, Xin-Bao Wu, Le-Wei Li, and Bo-Ran Guan.
\newblock Mie scattering by a uniaxial anisotropic sphere.
\newblock {\em physical review E}, 70(5):056609, 2004.

\bibitem{walker1957magnetostatic}
Laurence~R Walker.
\newblock Magnetostatic modes in ferromagnetic resonance.
\newblock {\em Physical Review}, 105(2):390, 1957.

\bibitem{sun2015hamiltonian}
Zhiyuan Sun, {\'A}~Guti{\'e}rrez-Rubio, DN~Basov, and MM~Fogler.
\newblock Hamiltonian optics of hyperbolic polaritons in nanogranules.
\newblock {\em Nano letters}, 15(7):4455--4460, 2015.

\bibitem{alvarez2019analytical}
Gonzalo {\'A}lvarez-P{\'e}rez, Kirill~V Voronin, Valentyn~S Volkov, Pablo
  Alonso-Gonz{\'a}lez, and Alexey~Y Nikitin.
\newblock Analytical approximations for the dispersion of electromagnetic modes
  in slabs of biaxial crystals.
\newblock {\em Physical Review B}, 100(23):235408, 2019.

\bibitem{valagiannopoulos2007study}
Constantinos Valagiannopoulos.
\newblock Study of an electrically anisotropic cylinder excited magnetically by
  a straight strip line.
\newblock {\em Progress In Electromagnetics Research}, 73:297--325, 2007.

\bibitem{lakhtakia2021theory}
Akhlesh Lakhtakia, Nikolaos~L Tsitsas, and Hamad~M Alkhoori.
\newblock Theory of perturbation of electrostatic field by an anisotropic
  dielectric sphere.
\newblock {\em The Quarterly Journal of Mechanics and Applied Mathematics},
  74(4):467--490, 2021.

\bibitem{lakhtakia1991electromagnetic}
Akhlesh Lakhtakia.
\newblock Electromagnetic response of an electrically small bianisotropic
  ellipsoid immersed in a chiral fluid.
\newblock {\em Berichte der Bunsengesellschaft f{\"u}r physikalische Chemie},
  95(5):574--576, 1991.

\bibitem{qiu2007scattering}
Cheng-Wei Qiu, Le-Wei Li, Tat-Soon Yeo, and Said Zouhdi.
\newblock Scattering by rotationally symmetric anisotropic spheres: Potential
  formulation and parametric studies.
\newblock {\em Physical Review E-Statistical, Nonlinear, and Soft Matter
  Physics}, 75(2):026609, 2007.

\bibitem{wallen2015anomalous}
Henrik Wall{\'e}n, Henrik Kettunen, and Ari Sihvola.
\newblock Anomalous absorption, plasmonic resonances, and invisibility of
  radially anisotropic spheres.
\newblock {\em Radio Science}, 50(1):18--28, 2015.

\bibitem{foteinopoulou2019phonon}
Stavroula Foteinopoulou, Ganga Chinna~Rao Devarapu, Ganapathi~S Subramania,
  Sanjay Krishna, and Daniel Wasserman.
\newblock Phonon-polaritonics: enabling powerful capabilities for infrared
  photonics.
\newblock {\em Nanophotonics}, 8(12):2129--2175, 2019.

\bibitem{beitner2024localized}
Daniel Beitner, Asaf Farhi, Ravindra~Kumar Nitharwal, Tejendra Dixit, Tzvia
  Beitner, Shachar Richter, SivaRama Krishnan, and Haim Suchowski.
\newblock Observation of localized resonant phonon polaritons in biaxial
  nanoparticles.
\newblock {\em Advanced Science}, 12(43), 2025.

\bibitem{kurs2007wireless}
Andre Kurs, Aristeidis Karalis, Robert Moffatt, John~D Joannopoulos, Peter
  Fisher, and Marin Soljacic.
\newblock Wireless power transfer via strongly coupled magnetic resonances.
\newblock {\em Science}, 317(5834):83--86, 2007.

\bibitem{li2003self}
Kuiru Li, Mark~I Stockman, and David~J Bergman.
\newblock Self-similar chain of metal nanospheres as an efficient nanolens.
\newblock {\em Physical review letters}, 91(22):227402, 2003.

\bibitem{dicke1954coherence}
Robert~H Dicke.
\newblock Coherence in spontaneous radiation processes.
\newblock {\em Physical review}, 93(1):99, 1954.

\bibitem{griffin2018directional}
Sinead Griffin, Simon Knapen, Tongyan Lin, and Kathryn~M Zurek.
\newblock Directional detection of light dark matter with polar materials.
\newblock {\em Physical Review D}, 98(11):115034, 2018.

\bibitem{curto2010unidirectional}
Alberto~G Curto, Giorgio Volpe, Tim~H Taminiau, Mark~P Kreuzer, Romain Quidant,
  and Niek~F Van~Hulst.
\newblock Unidirectional emission of a quantum dot coupled to a nanoantenna.
\newblock {\em Science}, 329(5994):930--933, 2010.

\bibitem{hillenbrand2002phonon}
R~Hillenbrand, T~Taubner, and F~Keilmann.
\newblock Phonon-enhanced light--matter interaction at the nanometre scale.
\newblock {\em Nature}, 418(6894):159--162, 2002.

\bibitem{pyatenko2007synthesis}
Alexander Pyatenko, Munehiro Yamaguchi, and Masaaki Suzuki.
\newblock Synthesis of spherical silver nanoparticles with controllable sizes
  in aqueous solutions.
\newblock {\em The Journal of Physical Chemistry C}, 111(22):7910--7917, 2007.

\bibitem{galanzha2017spaser}
Ekaterina~I Galanzha, Robert Weingold, Dmitry~A Nedosekin, Mustafa
  Sarimollaoglu, Jacqueline Nolan, Walter Harrington, Alexander~S Kuchyanov,
  Roman~G Parkhomenko, Fumiya Watanabe, Zeid Nima, et~al.
\newblock Spaser as a biological probe.
\newblock {\em Nature communications}, 8(1):15528, 2017.

\bibitem{okamoto2014fabrication}
Shinya Okamoto, Kazuhiro Inaba, Takuya Iida, Hajime Ishihara, Satoshi Ichikawa,
  and Masaaki Ashida.
\newblock Fabrication of single-crystalline microspheres with high sphericity
  from anisotropic materials.
\newblock {\em Scientific Reports}, 4(1):5186, 2014.

\bibitem{lakhtakia2021corrigendum}
Akhlesh Lakhtakia, Hamad~M Alkhoori, and Nikolaos~L Tsitsas.
\newblock Theory of perturbation of electric potential by a 3d object made of
  an anisotropic dielectric material (2021 j. phys. commun. 5 115010).
\newblock {\em Journal of Physics Communications}, 5(12), 2021.

\bibitem{farhi2021coupling}
Asaf Farhi and Aristide Dogariu.
\newblock Coupling of electrodynamic fields to vibrational modes in helical
  structures.
\newblock {\em Physical Review A}, 103(2):023523, 2021.

\bibitem{guzatov2010plasmon}
DV~Guzatov, VV~Klimov, and M~Yu Pikhota.
\newblock Plasmon oscillations in ellipsoid nanoparticles: Beyond dipole
  approximation.
\newblock {\em Laser Physics}, 20:85--99, 2010.

\bibitem{niven1891vi}
William~Davidson Niven.
\newblock Vi. on ellipsoidal harmonics.
\newblock {\em Philosophical Transactions of the Royal Society of London.(A.)},
  (182):231--278, 1891.

\bibitem{whittaker2020course}
Edmund~Taylor Whittaker and George~Neville Watson.
\newblock {\em A course of modern analysis}.
\newblock Courier Dover Publications, 2020.

\bibitem{hobson1931theory}
Ernest~William Hobson.
\newblock {\em The theory of spherical and ellipsoidal harmonics}.
\newblock CUP Archive, 1931.

\bibitem{carminati1999near}
R{\'e}mi Carminati and Jean-Jacques Greffet.
\newblock Near-field effects in spatial coherence of thermal sources.
\newblock {\em Physical Review Letters}, 82(8):1660, 1999.

\bibitem{biehs2021near}
S-A Biehs, Riccardo Messina, Prashanth~S Venkataram, Alejandro~W Rodriguez,
  Juan~Carlos Cuevas, and Philippe Ben-Abdallah.
\newblock Near-field radiative heat transfer in many-body systems.
\newblock {\em Reviews of Modern Physics}, 93(2):025009, 2021.

\bibitem{purcell1995spontaneous}
Edward~Mills Purcell.
\newblock Spontaneous emission probabilities at radio frequencies.
\newblock In {\em Confined electrons and photons: new physics and
  applications}, pages 839--839. Springer, 1995.

\bibitem{dung2002intermolecular}
Ho~Trung Dung, Ludwig Kn{\"o}ll, and Dirk-Gunnar Welsch.
\newblock Intermolecular energy transfer in the presence of dispersing and
  absorbing media.
\newblock {\em Physical Review A}, 65(4):043813, 2002purcell1995spontaneous,.

\bibitem{lee2024super}
Seungah Lee, Junghwa Lee, and Seong~Ho Kang.
\newblock Super-resolution multispectral imaging nanoimmunosensor for
  simultaneous detection of diverse early cancer biomarkers.
\newblock {\em ACS sensors}, 2024.

\bibitem{kadribasic2018directional}
Fedja Kadribasic, Nader Mirabolfathi, Kai Nordlund, Andrea~E Sand, Eero
  Holmstr{\"o}m, and Flyura Djurabekova.
\newblock Directional sensitivity in light-mass dark matter searches with
  single-electron-resolution ionization detectors.
\newblock {\em Physical Review Letters}, 120(11):111301, 2018.

\bibitem{boyd2023directional}
Christian Boyd, Yonit Hochberg, Yonatan Kahn, Eric~David Kramer, Noah Kurinsky,
  Benjamin~V Lehmann, and To~Chin Yu.
\newblock Directional detection of dark matter with anisotropic response
  functions.
\newblock {\em Physical Review D}, 108(1):015015, 2023.

\end{thebibliography}

\clearpage
\begin{widetext}
\section*{Supplementary Material}

\subsection{Full-wave simulations of uniaxial and biaxial particles}
\begin{figure}[ht]
\includegraphics[width=9cm]{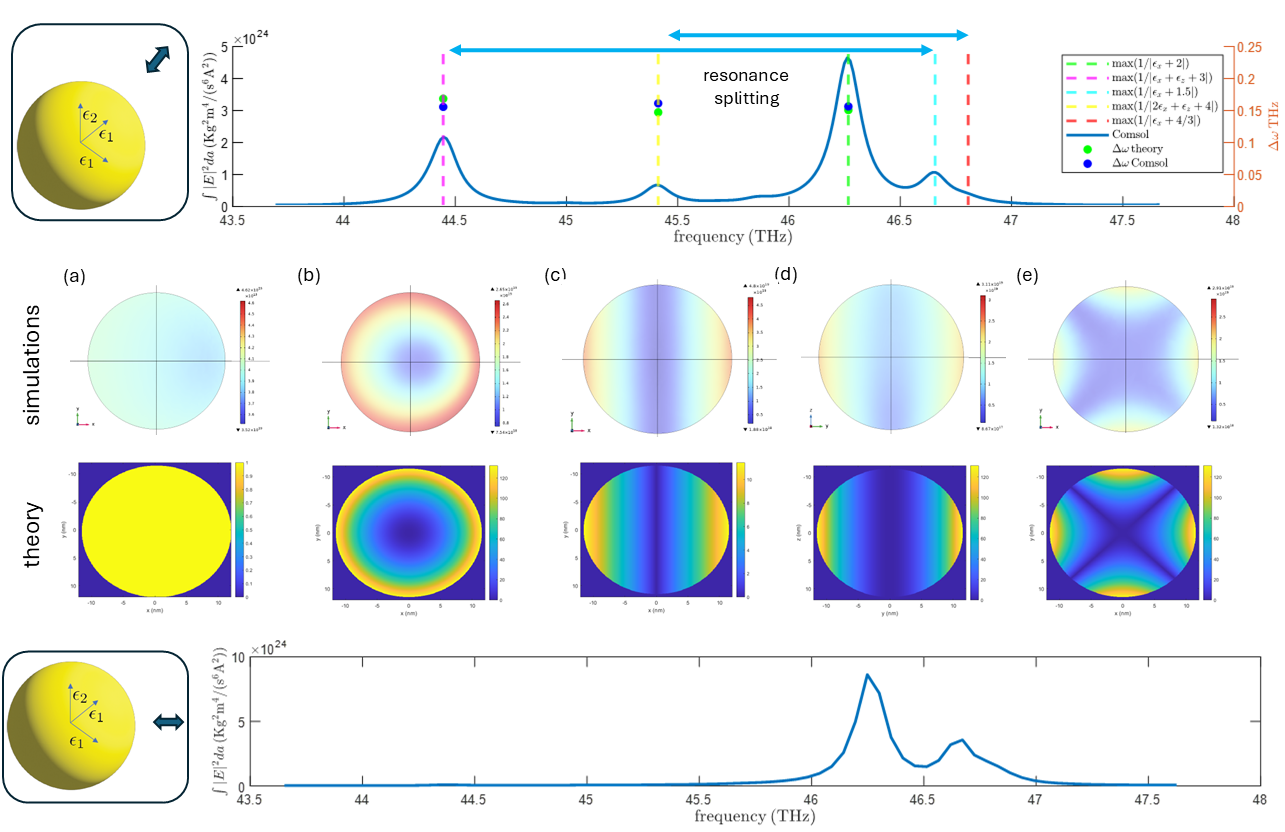}
\caption{Scattering spectrum, peak widths, and fields near resonance for a uniaxial hBN sphere of a radius $a=11.5\,\mathrm{nm}$ excited by an oscillating dipole located at $\mathbf{r}_0=(1,0,1)\cdot22\,\mathrm{nm}$ with a dipole moment of $\mathbf{P}=(1,0,1)\,\mathrm{A\cdot m}.$. Top: $\int |\mathbf{E}|^2 da$ over the particle envelope  calculated in COMSOL and compared to the peaks predicted analytically using $1/|\mathrm{resonance\,condition}|$ with excellent agreement. The circles are the peak widths obtained analytically from $1/|\mathrm{resonance\,condition}(\omega)|^2=\mathrm{max}/2$ compared to the ones calculated from the simulations, with very good agreement. (a),(b),(c),(d),(e) $|\mathbf{E}|$ for the  resonance conditions: $\epsilon_{1x}=-2,$ $\epsilon_{1x}=-1.5,$ $\epsilon_{1x}+\epsilon_{1z}=-3,$ $\epsilon_{1x}=-4/3,$ $2\epsilon_{1x}+\epsilon_{1x}=-4$ calculated at $f=46.301,\,$ $46.619,$ $\,44.53,\,$ $\,46.725,$ $45.401\mathrm{(THz)},$
respectively, using COMSOL compared to the  mode fields calculated analytically with very good agreement. Bottom: $\int |\mathbf{E}|^2 da$ over the particle envelope calculated in COMSOL for a dipole located at $\mathbf{r}_0=(1,0,0)31\mathrm{nm}$ (same dipole distance as the previous case) with a dipole moment of $\mathbf{P}=(1,0,0)\,\mathrm{A\cdot m}.$ As predicted the $m=\pm(l-1)$ modes disappeared from the spectrum.}
\end{figure}

\begin{figure}[htbp]
\centering
\begin{tabular}{cc}
    \includegraphics[width=4cm]{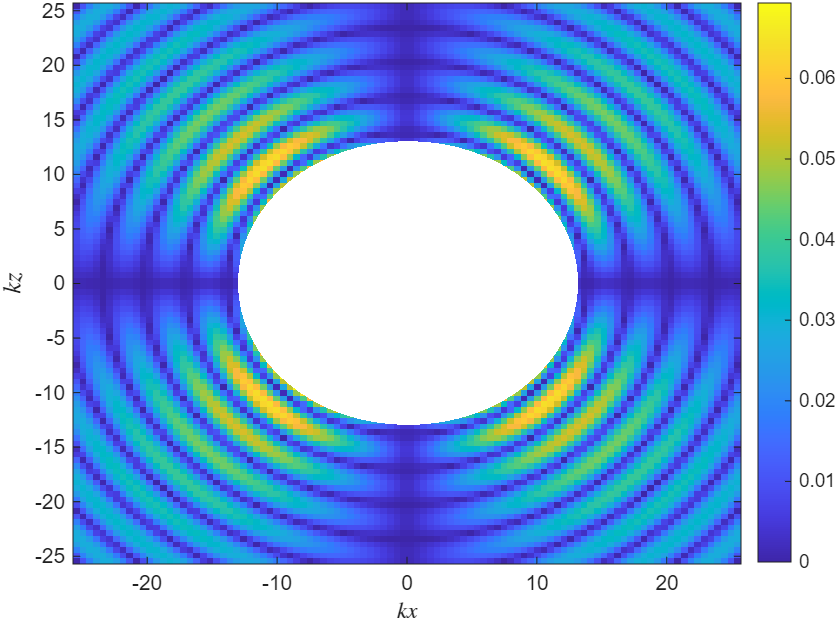} &
    \includegraphics[width=4cm]{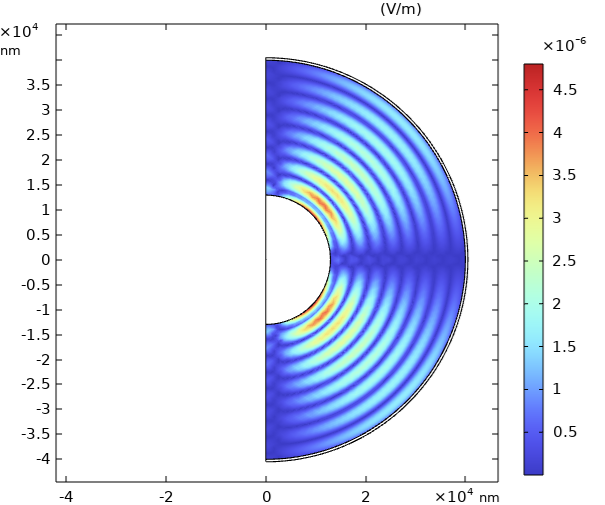} \\
     (a)  & (b) 
\end{tabular}
\caption{Radiation pattern calculations for a uniaxial particle excited similarly to Fig. 1 (top). (a) Closed-form radiation-pattern plot of $|\mathbf{E}|$  corresponding to  $\psi_{2}^{2}+\psi_{2}^{-2}$ outside the particle. (b) Full-wave COMSOL simulation of the scattered field for a uniaxial particle close to a resonance with $\epsilon_{1x}=-1.49$, $\epsilon_{1y}=-1.49$, $\epsilon_{1z}=1$ in $r>2\lambda$.}
\label{fig:two_subfigs}
\end{figure}


\begin{figure}
\includegraphics[width=9cm]{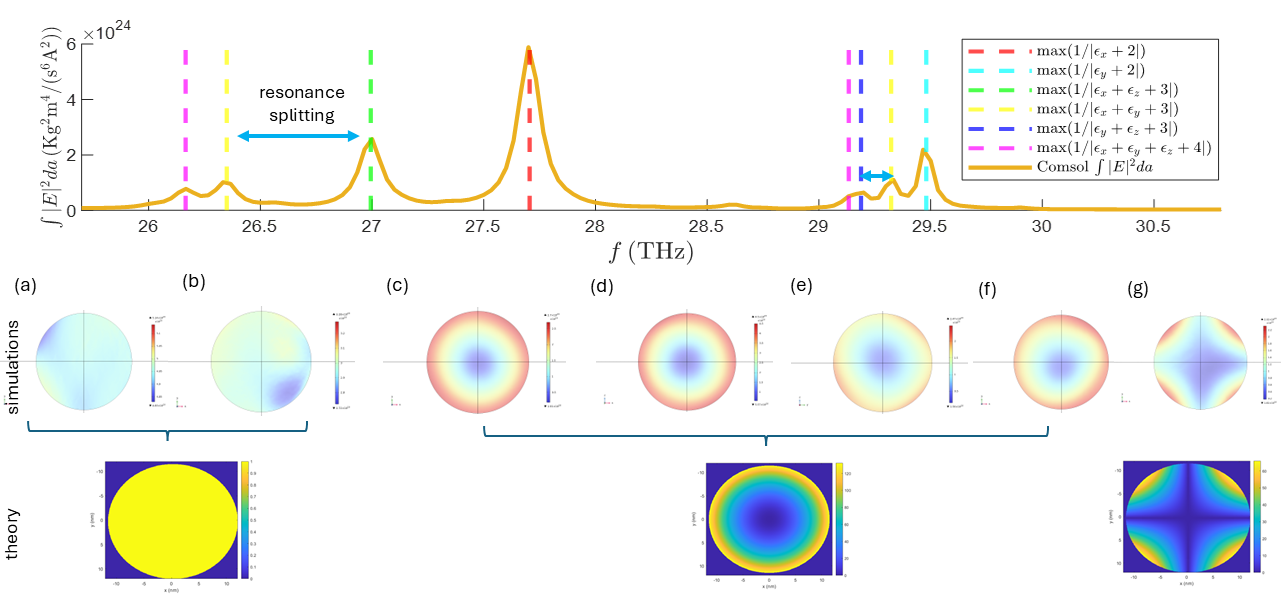}
\caption{Scattering spectrum and modes/fields near resonance for a biaxial $\alpha$-MoO3 sphere of a radius $a=11.5\,\mathrm{nm}$ excited by an oscillating dipole located at $\mathbf{r}_0=(18,18,18)\,\mathrm{nm}$ with a dipole moment of $\mathbf{P}=(1,1,1)\,\mathrm{A\cdot m}.$. Top:  $\int |\mathbf{E}|^2 da$ calculated in COMSOL and compared to the peaks predicted analytically using $1/|\mathrm{resonance\,condition}|$ with excellent agreement. (a),(b),(c),(d),(e),(f),(g) $|\mathbf{E}|$ for the resonance conditions: $\epsilon_{1x}=-2,$ $\epsilon_{1y}=-2,$ $\epsilon_{1x}+\epsilon_{1y}=-3,$ $\epsilon_{1x}+\epsilon_{1z}=-3,$ $\epsilon_{1y}+\epsilon_{1z}=-3,$ $\epsilon_{1x}+\epsilon_{1y}=-3,$ $\epsilon_{1x}+\epsilon_{1y}+\epsilon_{1z}=-4,$ calculated at $f=27.7, 29.7, 26.6,27,29.2, 29.33,26.167\,\mathrm{(THz)}$ respectively, using COMSOL compared to the  mode fields calculated analytically  with very good agreement. }
\end{figure}
\begin{figure*}[t]
\includegraphics[width=15cm]{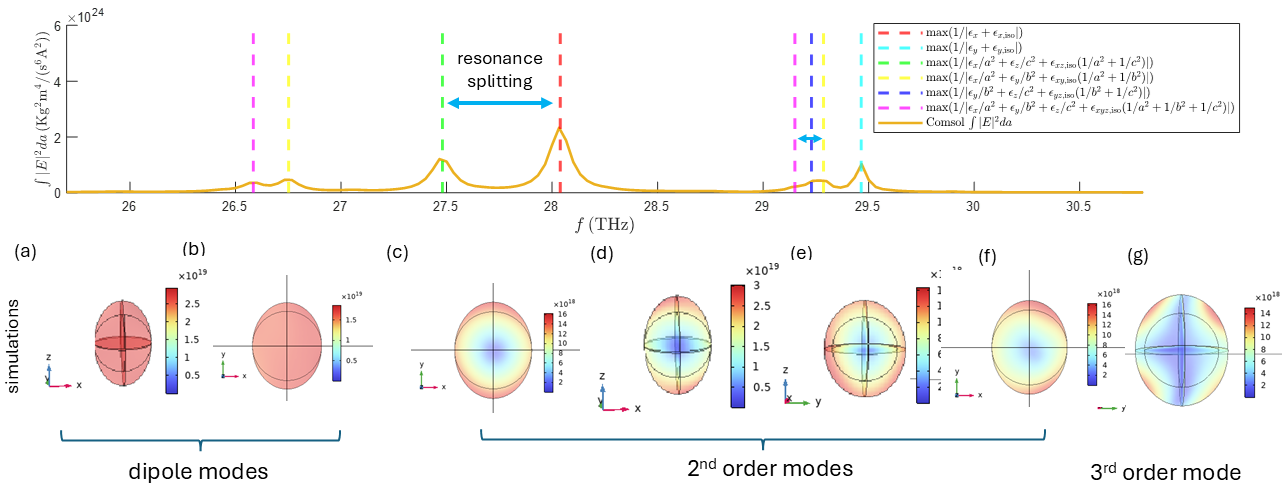}
\caption{Scattering spectrum and modes/fields near resonance for a biaxial $\alpha$-MoO3 ellipsoid with semiaxes $(a,b,c)=(1,1.25,1.5)\cdot 11.5\,\mathrm{nm}$ excited by an oscillating dipole located at $\mathbf{r}_0=(22,22,22)\,\mathrm{nm}$ with a dipole moment of $\mathbf{P}=(1,1,1)\,\mathrm{A\cdot m}.$. Top: Surface integral over the particle envelope $\int |\mathbf{E}|^2 da$ calculated in COMSOL and compared to the peaks predicted analytically using $1/|\mathrm{resonance\,condition}|$ with excellent agreement. (a),(b),(c),(d),(e),(f),(g) $|\mathbf{E}|$ for the resonance conditions: $\epsilon_{1x}=\epsilon_{1x,\mathrm{iso}},$ $\epsilon_{1y}=\epsilon_{1y,\mathrm{iso}},$ $\epsilon_{1x}a^{-2}+\epsilon_{1y}b^{-2}=\epsilon_{1xy,\mathrm{iso}}(a^{-2}+b^{-2}),$ $\epsilon_{1x}a^{-2}+\epsilon_{1z}c^{-2}=\epsilon_{1xz,\mathrm{iso}}(a^{-2}+c^{-2}),$ $\epsilon_{1y}b^{-2}+\epsilon_{1z}c^{-2}=\epsilon_{1yz,\mathrm{iso}}(b^{-2}+c^{-2}),$ $\epsilon_{1x}a^{-2}+\epsilon_{1y}b^{-2}=\epsilon_{1xy,\mathrm{iso}}(a^{-2}+b^{-2}),$ $\epsilon_{1x}a^{-2}+\epsilon_{1y}b^{-2}+\epsilon_{1z}c^{-2}=\epsilon_{1xyz,\mathrm{iso}}(a^{-2}+b^{-2}+c^{-2}),$ calculated at $f=28.033, 29.467, 26.6733,27.467,29.233, 29.3,26.6\,\mathrm{(THz)}$ respectively, using COMSOL with very good agreement to the mode fields calculated analytically (same as in  Fig. 3, defined in the ellipsoid volume). }
\end{figure*}
We demonstrate our theory for spheres and ellipsoids composed of hBN and $\alpha-\mathrm{MoO_3}$ in full-wave electrodynamic simulations. \hl{While our theory applies to any uniaxial and biaxial material, we have chosen these materials as they are currently at the forefront of photonics research and enable to achieve high Q factors. Whereas calculating  the resonance conditions, modes, Q factors, and radiation patterns from our theory is immediate, the full-wave hyperspectral simulations require substantial computational effort, taking approximately 7, 4, and 7 hours for the uniaxial particle, biaxial particle, and biaxial ellipsoid, respectively, where the biaxial particle simulations had coarse frequency sampling and the biaxial-particle frequency sampling included only two out of three bulk resonances. In particular, biaxial particles and localized sources cannot be simulated utilizing 2D axial symmetry and instead require full 3D simulations over numerous frequencies. Because localized excitation is common in high-order mode excitation and near-field experiments, our approach provides a crucial computational advantage, as it enables tuning of ellipsoidal particles to a target frequency by scanning the entire ellipsoid aspect-ratio parameter space, with applications in sensing and imaging. In contrast, hyperspectral sampling of the entire anistropic ellispoid aspect-ratio parameter space with simulations is computationally prohibitive.}

We consider a subwavelength uniaxial hBN sphere with the equal permittivity axes $x,y$ and the permittivity model used in Figs. 3,4 in the main text  \cite{foteinopoulou2019phonon}. The sphere has a radius of $11.5\mathrm{nm},$ and we situate an oscillating dipole at  $\mathbf{r}_0=(22\,\mathrm{nm},0,22\,\mathrm{nm})$ with a dipole moment of $\mathbf{P}=(1,0,1)\,\mathrm{A\cdot m}.$ We perform frequency domain simulations with a frequency sweep from $f=4.3652\cdot 10^{13}\,$(1/s) to $f=4.7626\cdot 10^{13}\,$(1/s) with 304 frequencies using extensive COMSOL simulations. The dipole moment was chosen to be in the direction of $\hat{r}$ since we assume \cite{farhi2021coupling,farhi2016electromagnetic,farhi2017eigenstate}, and prove in the SM, that the mode excitation coefficient depends on $\nabla\psi_l(\mathbf{r}_0)\cdot \mathbf{p}$ so that we need to incorporate only $\partial\psi_l/\partial r,$ preserving the angular dependence in  $\nabla\psi_l(\mathbf{r}_0)\cdot \mathbf{p}$. This, together with the fact that the dipole has only $x,z$ components, ensures that the degenerate modes will add constructively, as is explained in the SM. Note that the coefficient's dependence on the mode fields at the dipole location, with the rapid decay of the high-order modes away from the particle, reflect the presence of higher spatial frequencies near the dipole, which excite higher-order modes. \hl{For a general incoming field, the mode excitation coefficient is given by  $\left\langle \psi_{k}|\psi_{0}\right\rangle =\int\theta_1\nabla\psi_{k}^{*}\cdot\nabla\psi_{0}d\mathbf{r'},$ where $\theta_1$ is a window function that equals one in the particle volume,  and we utilized a dipole source since a far field  excites high-order modes with negligible magnitudes}. We then calculate the resonance peak widths analytically (from 1/resonance condition) and from the simulations. To demonstrate the mode-coefficient dependency on the dipole location and orientation, we also perform the same calculation for a dipole oriented along the $x$ axis. In this case, we do not expect the $m=\pm(l-1)$ modes to show in the spectrum since in order to excite them a $z$ component of the dipole location is required.
In Fig. 1 (top panel) we present the analytically-predicted peaks in the scattering spectrum obtained from the maxima of $1/|\mathrm{resonance}\, \mathrm{condition}|$ compared to the surface integral of $|\mathbf{E}|^2$ over the uniaxial particle envelope calculated in COMSOL with excellent agreement. Importantly, these simulation results for the scattering spectrum confirm our predictions of the resonance splitting (marked by arrows) and mode degeneracy for uniaxial spheres. We also obtained the peak widths analytically from $1/|\mathrm{resonance\,condition}(\omega)|^2=\mathrm{max}/2$ and compared them to the ones calculated from the simulations (as can be seen, the width of the peak, which corresponds to $\epsilon_x=-1.5,$ is altered due to the peak on the right, and therefore we haven't analyzed it). Clearly, there is remarkable agreement between them, which  showcases our important ability to analytically predict the Q factors given by $\omega/\Delta \omega.$
We then plot for all the observed peaks $|\mathbf{E}|$ inside the particle calculated in COMSOL and obtained analytically from the modes (Methods). While the COMSOL fields include the incoming field from the oscillating dipole, close to a resonance this is typically negligible and indeed we observe very good agreement. Moreover, we plot $\int|\mathbf{E}|^2 da$ for a dipole located at the same distance as the previous case  and oriented along $x$ ($\mathbf{r}_0=(1,0,0)$31nm). As we predicted the $m=\pm(l-1)$ modes disappeared from the spectrum, verifying the dependency of the mode excitation coefficient on the dipole location and orientation. This highlights the directional nature of the modes as opposed to isotropic particles. \hl{In Fig. 2 we compared the radiation pattern from the full-wave closed-form solution corresponding to the mode $\psi_{2}^{2}+\psi_{2}^{-2}$ with the resonance condition  $\epsilon_{1x}=\epsilon_{1y}=-1.5$ outside a uniaxial particle to a full-wave COMSOL simulation of the scattered field of a uniaxial particle with $R=46$ nm and $\epsilon_{1x}=-1.49,\epsilon_{1y}=-1.49,\epsilon_{1z}=1,$  excited by a current source similarly to Fig. 1 (top) \emph{but rotated axially (occupying a ring)} in the radiation zone $(r>2\lambda)$. Note that calculating the radiation patterns is computationally more demanding due to the large simulation volume. Therefore, only for this case, we employed 2D axially symmetric simulations, which impose an artificial ring-like source at a single frequency identified from the 3D simulations of the uniaxial particle. As can be seen, the radiation-pattern results show very good agreement.}


We then considered a biaxial subwavelength sphere composed of $\alpha-$MoO3 with the permittivity model in Ref. \cite{foteinopoulou2019phonon}. The sphere is excited by a dipole located at $\mathbf{r}_0=(18\,\mathrm{nm},18\,\mathrm{nm},18\,\mathrm{nm})$ with the dipole moment of $\mathbf{P}=(1,1,1) \,\mathrm{A\cdot m}.$
We performed frequency domain simulations with a frequency sweep between $f=25.5\,\mathrm{(THz)}$ and $f=30.8\,\mathrm{(THz)}$ with 175 frequencies using COMSOL. Here, again, the dipole moment is directed along $\hat{r},$ to preserve the angular dependency of the mode in the mode-excitation coefficient. In Fig. 3 we present the result of the surface integral $\int |\mathbf{E}|^2da$ over the particle envelope from the COMSOL simulations and the predicted peaks from the resonance conditions with very good agreement. The predicted peaks were readily calculated by finding the maxima of $1/|\mathrm{resonance}\,\mathrm{condition}|$. Crucially, our simulations confirm the resonance splitting effect predicted analytically, see arrows in Fig. 3. We then compared for all the observed resonances $|\mathbf{E}|$ from the simulations to the ones obtained analytically from the modes with excellent agreement. We also report very good agreement between the peak frequency widths obtained analytically and the simulations (accurate comparison would require a highly-dense frequency sweep and simulations that are very extensive).

Finally, we simulated a biaxial subwavelength ellipsoid of the most general geometry with semiaxes $(a,b,c)=(1,1.25,1.5)\cdot 11.5\,\mathrm{nm},$  composed of $\alpha-$MoO3 with the permittivity model in Ref. \cite{foteinopoulou2019phonon}. The ellipsoid is excited by a dipole located at $\mathbf{r}_0=(22\,\mathrm{nm},22\,\mathrm{nm},22\,\mathrm{nm})$ with the dipole moment of $\mathbf{P}=(1,1,1) \,\mathrm{A\cdot m}.$
We performed frequency domain simulations with a frequency sweep between $f=25.5\,\mathrm{(THz)}$ and $f=30.8\,\mathrm{(THz)}$ with 175 frequencies using COMSOL. In Fig. 4 we present the result of the surface integral $\int |\mathbf{E}|^2da$ over the particle envelope from the COMSOL simulations and the predicted peaks from the resonance conditions with very good agreement. The predicted peaks were readily calculated by finding the maxima of $1/|\mathrm{resonance}\,\mathrm{condition}|$.  We then compared for all the observed resonances $|\mathbf{E}|$ from the simulations to the ones obtained analytically  from the modes (equivalent to Fig. 3) with excellent agreement. We note that the isotropic-ellipsoid eigenpermittivities that were calculated to obtain the biaxial ellipsoid resonance permittivity relations are the following: $\epsilon_{\mathrm{iso}\,x}= -1.387,$
$\epsilon_{\mathrm{iso}\,y}= -2.092,$
$\epsilon_{\mathrm{iso}\,xy}= -1.295,$
$\epsilon_{\mathrm{iso}\,yz}= -1.748,$
$\epsilon_{\mathrm{iso}\,xz}= -1.385,$
$\epsilon_{\mathrm{iso}\,xyz}= -1.287.$ 
As can be seen, for an isotropic ellipsoid with these semi axis ratios $\epsilon_{\mathrm{iso}\,x}\approx \epsilon_{\mathrm{iso}\,xz}$ and $\epsilon_{\mathrm{iso}\,xy}\approx\epsilon_{\mathrm{iso}\,xyz},$ and due to the anisotropic material these resonances are split, as can be seen in Fig. 4. In contrast to spheres, where the degeneracy is enforced by symmetry, here the degeneracy is accidental.
Finally, there is a clear physical intuition for the resonance conditions of anisotropic ellipsoids: the resonances of isotropic ellipsoids, on which they are based, tend to favor shorter ellipsoid axes, since they exhibit higher real permittivity values that have smaller imaginary parts of the physical permittivity,  leading to stronger and sharper resonances.

\subsection{Superimposing the degenerate uniaxial modes for the sphere simulation comparison}

\subsubsection{Theory}

We now analytically derive the field distributions and resonance frequencies for the uniaxial sphere and the dipole location in the main text.
\subsubsection*{Dipole mode}
The resonance condition is $\epsilon_{x}+2=0$ and we therefore find the maximum of $1/|\epsilon_{x}+2|,$
which occurs for hBN \cite{foteinopoulou2019phonon} at $f=4.6265\cdot10^{13}\left(\frac{1}{s}\right).$   The mode field inside the particle is given by
$\mathbf{E}_{\mathrm{ins}}=\hat{x}.$

\subsubsection*{2nd order modes}
For the first 2nd-order mode we get from the theory
$\epsilon_{1x}+1.5=0,$ which for hBN approximately occurs at $f=4.6655\cdot10^{13}\left(\frac{1}{s}\right).$ We add the degenerate modes, which are excited with equal strengths:
\begin{gather}
\psi_{22}^{\pm}=\left(x\pm iy\right)^{2},\,\,
\psi_{22}^{\pm}=x^{2}\pm ixy-y^{2},
\nonumber\\
\psi_{22}^{+}+\psi_{22}^{-}\propto x^{2}-y^{2},\nonumber\\
\boldsymbol{E}_{22}^{+}+\boldsymbol{E}_{22}^{-}=2x\hat{x}-2y\hat{y},\,\,
\left|\boldsymbol{E}_{22}^{+}+\boldsymbol{E}_{22}^{-}\right|\propto\sqrt{x^{2}+y^{2}}.\nonumber
\end{gather}
\paragraph{second mode}

The derived resonance condition reads
\[
\epsilon_{1x}+\epsilon_{1z}+3=0.
\]
For hBN this is approximately satisfied for $f=4.453\cdot10^{13}\left(\mathrm{\frac{1}{s}}\right).$ 
Now we calculate the field by adding the degenerate modes:
\begin{gather}
\psi_{21}^{\pm}=\left(x\pm iy\right)z,\,\,
\psi_{21}^{+}+\psi_{21}^{-}\propto xz,\nonumber\\
\boldsymbol{E}_{21}^{+}+\boldsymbol{E}_{21}^{-}=z\hat{x}+x\hat{z},\,\,
\left|\boldsymbol{E}_{21}^{+}+\boldsymbol{E}_{21}^{-}\right|=\sqrt{z^{2}+x^{2}},
\nonumber\\
\left|\boldsymbol{E}_{21}^{+}\left(z=0\right)+\boldsymbol{E}_{21}^{-}\left(z=0\right)\right|=\left|x\right|.\nonumber
\end{gather}

\subsubsection*{3rd order modes}

\paragraph{First mode}
The resonance condition reads:
\[
\epsilon_{1x}+4/3=0.
\]
For hBN this occurs at $f=4.6798\cdot10^{13}\left(\frac{1}{s}\right)$
. Adding the degenerate modes we get
\begin{gather}
\psi^{\pm}=\left(x\pm iy\right)^{3},\nonumber\\
\psi^{\pm}=x^{3}\pm2ix^{2}y-y^{2}x\pm iyx^{2}-2xy^{2}\mp iy^{3},\nonumber\\
\mathbf{E}^{\pm}=\left(3x^{2}\pm4ixy-y^{2}\pm2iyx-2y^{2}\right)\hat{x}\nonumber\\
+\left(\pm2ix^{2}-2xy\pm ix^{2}-4xy\mp3iy^{2}\right)\hat{y}.\nonumber
\end{gather}
To simplify things, we look at the $yz$ plane
\begin{gather}
\mathbf{E}^{\pm}\left(x=0\right)=-3y^{2}\hat{x}+\left(\pm3iy^{2}\right)\hat{y},\nonumber\\
\mathbf{E}^{+}\left(x=0\right)+\mathbf{E}^{-}\left(x=0\right)\propto-3y^{2}\hat{x},
\nonumber\\
\left|\mathbf{E}^{+}\left(x=0\right)+\mathbf{E}^{-}\left(x=0\right)\right|\propto3y^{2}.\nonumber
\end{gather}

\paragraph{Second mode}

The resonance condition $2\epsilon_{1x}+\epsilon_{1x}+4=0$ is satisfied
for hBN for $f=4.5409\,\mathrm{\left(\frac{1}{s}\right)}.$ Adding the modes we obtain:
\begin{gather}
\psi_{31}^{\pm}=\left(x\pm iy\right)^{2}z,
\nonumber\\
\boldsymbol{E}_{31}^{\pm}=\left(x\pm iy\right)z\hat{x}\pm i\left(x\pm iy\right)z\hat{y}+\left(x\pm iy\right)z\hat{y},
\nonumber\\
\psi_{31}^{+}+\psi_{31}^{-}\propto\left(x^{2}-y^{2}\right)z,
\nonumber\\
\boldsymbol{E}_{31}^{+}+\boldsymbol{E}_{31}^{-}=2xz\hat{x}-2yz\hat{y}+\left(x^{2}-y^{2}\right)\hat{z},\nonumber\\
\left|\boldsymbol{E}_{31}^{+}+\boldsymbol{E}_{31}^{-}\right|=\sqrt{\left(2xz\right)^{2}+\left(2yz\right)^{2}+\left(x^{2}-y^{2}\right)^{2}},\nonumber\\
\left|\mathbf{E}\left(z=0\right)\right|=\left|x^{2}-y^{2}\right|.
\nonumber
\end{gather}
We plot these field magnitude distributions in Fig. 1.
One can express the eigenvalue in the denominator $s_{1xl}$ in terms of the physical parameter $\epsilon_{1z}$ using the eigenpermittivity relations, obtaining exactly the resonance conditions now for the physical parameters used along the paper (e.g., ones used in Figs. 3 and 4). Note that the first sum includes only the $z$ dipole term and the second sum includes the remaining eigenstates in agreement with our analysis along the paper. It is worth noting that this formalism should also apply to other types of anisotropy such as spherical.

\subsection{Correspondence between the derived modes and the radiation patterns}
Here we derive the correspondence between the localized modes that we obtained in the main text and the radiation patterns of the anisotropic subwavelength spheres. We utilize the TM vector spherical harmonics (VSH) modes for a general spherical particle given by $\mathbf{E}\propto \nabla\times f_{l}\left(r\right)\boldsymbol{{X}}_{lm}\left(\theta,\phi\right)$
where
$$f_{l}\left(r\right)=\left\{ \begin{array}{cc}
j_{l}\left(\sqrt{\epsilon_{1}}kr\right) & r<a\\
h_{l}^{\left(1\right)}\left(kr\right) & r>a
\end{array}\right.,$$
and $j_{l},h_{l}^{\left(1\right)}$ are the  spherical bessel function and spherical Hankel function of the first kind and $\boldsymbol{X}_{lm}=\left(\boldsymbol{r}\times\nabla\right)Y_{lm} $ \cite{bergman1980theory,farhi2020three}.  Calculating the eigenpermittivity from these modes in the limit of $a\ll \lambda$ gives the quasistatic eigenvalues. For example, the full electrodynamic mode gives the following eigenpermittivities for a sphere with $a=11.5$nm and $\lambda=600$nm : $\epsilon_{1,l=1}=-2.035 - 0.00357 i,\epsilon_{1,l=2}=-1.505 - 2.12\cdot 10^{-6}i,\epsilon_{1,l=3}=-1.335 - 7.292\cdot 10^{-10} i,$ which are very close to the quasistatic results of $\epsilon_{1l}=-(l+1)/l.$ Even though these modes are usually considered for isotropic spheres, the results can be applied to anisotropic spheres since their modes are superpositions of spherical harmonics, which correspond to superpositions of VSHs. We write a TM VSH outside the sphere: 
\begin{gather}
\nabla\times h_{l}^{\left(1\right)}\left(kr\right)\boldsymbol{X}_{lm}=\nabla h_{l}^{\left(1\right)}\times\boldsymbol{X}_{lm}+h_{l}^{\left(1\right)}\left(r\right)\nabla\times\boldsymbol{X}_{lm},\nonumber
\end{gather}
Using the vector identity for $\boldsymbol{a}\times\boldsymbol{b}\times\boldsymbol{c}=\mathbf{b}(\mathbf{a}\cdot \mathbf{c})-\mathbf{c}(\mathbf{a}\cdot\mathbf{ b})$
we get for the first term:
\[
\nabla h_{l}^{\left(1\right)}\times\boldsymbol{X}_{lm}=\nabla h_{l}^{\left(1\right)}\times\boldsymbol{r}\times\nabla Y_{lm}=-r\frac{\partial h_{l}^{\left(1\right)}(kr)}{\partial r}\nabla Y_{lm}.
\]
For the second term we arrive at 
\begin{gather}
h_{l}^{\left(1\right)}\left(kr\right)\nabla\times\boldsymbol{X}_{lm}=h_{l}^{\left(1\right)}\left(kr\right)\left(\boldsymbol{r}\nabla^{2}Y_{lm}-\nabla Y_{lm}\right),\nonumber
\end{gather}
where we have used the vector identity for $\nabla\times\left(\boldsymbol{A}\times\boldsymbol{B}\right),$ $\nabla\cdot\boldsymbol{r}=3,$ $(\nabla Y_{lm}\cdot\nabla)\mathbf{r}=\nabla Y_{lm},$ $\left(\boldsymbol{r}\cdot\nabla\right)\nabla Y_{lm}=-\nabla Y_{lm}.$
Then, we utilize the spherical-harmonic identity \cite{jackson2021classical}
\[
\nabla^{2}Y_{lm}=-l\left(l+1\right)/r^{2}Y_{lm},
\]
to obtain
\begin{gather}
h_{l}^{\left(1\right)}\left(kr\right)\nabla\times\boldsymbol{X}_{lm}
=h_{l}^{\left(1\right)}\left(kr\right)\left[-l\left(l+1\right)/rY_{lm}\hat{r}-\nabla Y_{lm}\right].\nonumber
\end{gather}
All in all, we get:
\begin{gather}
\nabla\times h_{l}^{\left(1\right)}\left(kr\right)\boldsymbol{X}_{lm}\nonumber\\
=-r\frac{\partial h_{l}^{\left(1\right)}(kr)}{\partial r}\nabla Y_{lm}+h_{l}^{\left(1\right)}\left(kr\right)\left(-\frac{l\left(l+1\right)}{r}Y_{lm}\hat{r}-\nabla Y_{lm}\right).\nonumber
\end{gather}
We see that 
 when $r\rightarrow0$ since $h_{l}^{\left(1\right)}\propto \frac{e^{ikr}}{r^{l+1}},$ $r\partial h_{l}^{\left(1\right)}/\partial r\propto \frac{e^{ikr}}{r^{l+1}}$ the overall scaling is of $1/r^{l+2},$ in agreement with the quasistatic analysis.  For $r\rightarrow\infty$ since   $\frac{\partial h_{l}^{(1)}}{\partial r}=\frac{ike^{ikr}}{r}-\frac{e^{ikr}}{r^{2}},\,
\underset{r\to\infty}{\mathrm{lim}}\frac{\partial h_{l}^{(1)}}{\partial r}=\frac{ike^{ikr}}{r}$ and $\underset{r\to\infty}{\mathrm{lim}}\frac{h_{l}^{(1)}}{r}=\frac{e^{ikr}}{r^2}$ 
the first term, which is transverse and has $1/r$ scaling as expected, dominates. 
Thus, one can conclude from the superposition of $Y_{lm}$ that participate in the
quasistatic anisotropic modes, that the radiation patterns are of the form:
\[
\mathbf{E}_\mathrm{rad}\approx\sum_{l',m'}-ike^{ikr}\nabla Y_{l'm'},
\]
where the sum is over the  mode orders which participate in the anisotropic-sphere modes.
Please note that $\nabla Y_{lm}\propto\frac{1}{r}$ which gives $1/r$
overall scaling of the radiated field and we get the following radiation pattern:
\[
\mathbf{E}_\mathrm{rad}\approx-\sum_{l',m'} \text{\ensuremath{\frac{ike^{ikr}}{r}\left(\mathbf{e}_\theta\frac{\partial}{\partial\theta},\mathbf{e}_\phi\frac{1}{\sin\theta}\frac{\partial}{\partial\phi}\right)Y_{l'm'}}}.
\]
One can readily see that the $l=1,m=0$ mode of a subwavelength  sphere gives the dipole radiation pattern of $\mathbf{E}_\mathrm{rad}=\frac {e^{ikr}}{r}\sin\theta\mathbf{e}_\theta$. In addition, the radiation patterns of high-order anisotropic particle modes differ from the isotropic case. Note that while the uniaxial-sphere modes are identical to the $l,l-1$ isotropic sphere modes, in nearly all practical scenarios, their radiation patterns  differ due to the fact that an excitation source will excite  distinct superpositions of modes in the two cases.

\subsection{Biaxial sphere modes as superpositions of spherical harmonics}
We express the second-order biaxial modes as superpositions of spherical harmonics:
\begin{gather}
\tilde{\psi}_{2,b1}=\left(\psi_{2}^{-1}+\psi_{2}^{1}\right)/2=xzf_l(r),\nonumber\\
\tilde{\psi}_{2,b2}=\left(\psi_{2}^{1}-\psi_{2}^{-1}\right)/2i=yzf_l(r),\nonumber\\
\tilde{\psi}_{2,b3}=\left(\psi_{2}^{2}-\psi_{2}^{-2}\right)/2i=xyf_l(r).\nonumber
\end{gather}
We now proceed to the third and fourth-order biaxial modes. We continue along the lines of the second-order biaxial mode and write $\tilde{\psi}_3=\frac{\psi^2_3-\psi^{-2}_3}{2i}\propto xyz,$ $\tilde{\psi}_{b4}=\psi_{4}^{2}-\psi_{4}^{-2}+\frac{\psi_{4}^{4}-\psi_{4}^{-4}}{\sqrt{7}\cdot4i}=xyz^{2}.$ Using the procedure above, it can be shown that $\epsilon_{1x}+\epsilon_{1y}+\epsilon_{1z}=-4$ and $\epsilon_{x}+\epsilon_{y}+2\epsilon_{z}=-5$ are the resonance conditions, respectively. Note that the modes $x^2yz,\,xy^2z$ and their resonance conditions readily follow.
\subsection{Completeness of the solution}
We now prove that the derived uniaxial-sphere modes are all the eigenmodes.
Let us assume that there is an additional eigenstate besides the four modes for each $l$.
Since the isotropic-sphere modes span the space inside the sphere, this eigenstate is a superposition of the remaining isotropic-sphere modes 
$$
\psi_{hl}=\sum_{l',|m'|<l-1}a_{l'm'} \psi_{l'}^{m'}.$$
An eigenstate must satisfy anisotropic Laplace's equation inside the inclusion, which reads:
$$\epsilon_{1x}\frac{\partial^2 \psi_{hl}}{\partial x^2}+\epsilon_{1x}\frac{\partial^2 \psi_{hl}}{\partial y^2}+\epsilon_{1z}\frac{\partial^2 \psi_{hl}}{\partial z^2}=0.$$ Since it is a superposition of isotropic-sphere modes we write:
\begin{gather}
\epsilon_{1x}\frac{\partial^2 \psi_{hl}}{\partial x^2}+\epsilon_{1x}\frac{\partial^2 \psi_{hl}}{\partial y^2}+
\left(\epsilon_{1x}+\epsilon_{1z}-\epsilon_{1x}\right)\frac{\partial^2 \psi_{hl}}{\partial z^2}=0,\nonumber\\
\left(\epsilon_{1z}-\epsilon_{1x}\right)\frac{\partial^2 \psi_{hl}}{\partial z^2}=0.
\end{gather}
However, all these modes have orders of $z$ that are 2 or higher (including from $r$) and each function is fundamentally different from the others. Hence, such an eigenstate does not exist, and we have a contradiction. Conclusion: the derived uniaxial sphere modes are all the modes. Note that the uniaxial mode basis does not span space as in the standard case.
\\
Similarly, we write for a biaxial particle 
\begin{gather}
\epsilon_{1x}\frac{\partial^2 \psi_{hl}}{\partial x^2}+\epsilon_{1y}\frac{\partial^2 \psi_{hl}}{\partial y^2}+\epsilon_{1z}\frac{\partial^2 \psi_{hl}}{\partial z^2}=0,\nonumber\\
\left(\epsilon_{1x}-\epsilon_{1z}\right)\frac{\partial^2 \psi_{hl}}{\partial x^2}+\left(\epsilon_{1y}-\epsilon_{1z}\right)\frac{\partial^2 \psi_{hl}}{\partial y^2}=0.\nonumber
\end{gather}
If $ \frac{\partial^2 \psi_{hl}}{\partial x^2}=-\frac{\partial^2 \psi_{hl}}{\partial y^2},$ we get $\epsilon_{1x}=\epsilon_{1y},$ which is a uniaxial medium, and we therefore exclude this case.
All the $|m|<l-1$ modes are isotropic (from the boundary conditions) and the $|m|\ge l-1$ modes satisfy $\epsilon_{1x}=\epsilon_{1y}$ when $ \frac{\partial^2 \psi_{hl}}{\partial x^2},\frac{\partial^2 \psi_{hl}}{\partial y^2}\neq 0.$ Even if such modes exist, their axial permittivities must satisfy both Laplace's equation (in a non-trivial manner) and the boundary condition, which imposes strong constraints, \hl{which the axial permittivities have to satisfy at a given frequency}. As a result, their potential contribution is typically weak.  Therefore, we expect that the dominant biaxial modes satisfy $\frac{\partial^2 \psi_{l}}{\partial x^2}=\frac{\partial^2 \psi_{l}}{\partial y^2}=0$ and considered  all such possible modes in the main text.

\subsection{Eigenmode expansion for an anisotropic inclusion}
Eigenstate expansions have been derived for an isotropic inclusion and for an inclusion with only one axial permittivity that differs from the host medium permittivity \cite{bergman1980theory,farhi2021coupling}.
Here we expand the quasielectrostatic potential for an anisotropic inclusion. 
For concreteness, we consider a uniaxial inclusion and the derived eigenstates. Generalizing it to a biaxial inclusion readily follows. 

We start by writing Poisson's equation for a medium with an anisotropic inclusion:
\begin{gather}
\nabla\overleftrightarrow{\epsilon}\nabla\psi=\rho,\nonumber\\
\nabla\left[\epsilon_{2}+\theta_{1}\left(\overleftrightarrow{\epsilon}_{1}-\epsilon_{2}I\right)\right]\nabla\psi=\rho,\nonumber\\
\epsilon_{2}\nabla^{2}\psi+\nabla\theta_{1}\left(\overleftrightarrow{\epsilon}_{1}-\epsilon_{2}I\right)\nabla\psi=\rho,\nonumber\\
\epsilon_{2}\nabla^{2}\psi=\nabla\theta_{1}\left(\epsilon_{2}I-\overleftrightarrow{\epsilon}_{1}\right)\left(\frac{\partial\psi}{\partial x},\frac{\partial\psi}{\partial y},\frac{\partial\psi}{\partial z}\right)+\rho,\nonumber\\
\epsilon_{2}\nabla^{2}\psi=\nabla\theta_{1}\left(\epsilon_{2}-\epsilon_{1x},\epsilon_{2}-\epsilon_{1y},\epsilon_{2}-\epsilon_{1z}\right)\left(\frac{\partial\psi}{\partial x},\frac{\partial\psi}{\partial y},\frac{\partial\psi}{\partial z}\right)+\rho,\nonumber\\
\nabla^{2}\psi=\nabla\theta_{1}\left(\frac{\epsilon_{2}-\epsilon_{1x}}{\epsilon_{2}}\frac{\partial\psi}{\partial x}+\left(\frac{\epsilon_{2}-\epsilon_{1y}}{\epsilon_{2}}\right)\frac{\partial\psi}{\partial y}+\frac{\epsilon_{2}-\epsilon_{1z}}{\epsilon_{2}}\frac{\partial\psi}{\partial z}\right)+\rho/\epsilon_{2}.\nonumber
\end{gather}
where $\theta_1$ is a step function that equals 1 inside the inclusion, $\overleftrightarrow{{\epsilon}_1}$ is the inclusion permittivity tensor, and $\epsilon_2$ is the host medium permittivity.
Now we move one term to the lhs in order to be able to define an
eigenvalue equation. 
\[
\nabla^{2}\psi-\nabla\theta_{1}u_{1z}\frac{\partial}{\partial z}\psi=u_{1x}\nabla\theta_{1}\left(\frac{\partial\psi}{\partial x}+\frac{\partial\psi}{\partial y}\right)+\rho/\epsilon_{2}.
\]
where $u_{1i}=\frac{\epsilon_2-\epsilon_{1i}}{\epsilon_2},\,\,s_{1i}=1/u_{1i}$. The Green function in this case will be for a medium with an inclusion
with anisotropy in the $z$ axis and axial permittivities that are equal to the host medium permittivity in the other axes. 
\begin{gather}
\nabla^{2}G_{2}-\nabla\theta_{1}u_{1z}\frac{\partial}{\partial z}G_{2}=\delta\left(\boldsymbol{r}-\boldsymbol{r}'\right),\nonumber\\
\tilde{\Gamma}\psi=u_{1x}\int G_{2}\nabla\theta_{1}\left(\frac{\partial\psi}{\partial x}\hat{x}+\frac{\partial\psi}{\partial y}\hat{y}\right)\nonumber\\
=-u_{1x}\int\theta_{1}\nabla\left(G_{2}\right)\cdot \left(\frac{\partial\psi}{\partial x}\hat{x}+\frac{\partial\psi}{\partial y}\hat{y}\right)\nonumber\\
=-u_{1x}\int\theta_{1}\left(\frac{\partial G_{2}}{\partial x},\frac{\partial G_{2}}{\partial y}\right)\cdot \nabla\psi=\Gamma_{xy}\psi\nonumber
\end{gather}
where we defined $\Gamma_{xy}$ to adjust it to an eigenvalue equation with an operator operating on $\psi$ and we performed integration by parts.

We express the electric potential and expand it using the anisotropic
eigenfunctions defined by Laplace's equation. In the following $\tilde{\psi}_0$ is defined for the particle with anisotropy in the $z$ axis. We write:
\[
\psi=u_{1x}\tilde{\Gamma}\psi+\tilde{\psi}_{0},\,\,\,
\psi=u_{1x}\Gamma_{xy}\psi+\tilde{\psi}_{0},
\]
\[
\left(1-u_{1x}\Gamma_{xy}\right)\psi=\tilde{\psi}_{0},
\]
\[
\psi=\frac{1}{1-u_{1x}\Gamma_{xy}}\tilde{\psi}_{0},
\]
\[
\psi=\frac{s_{1x}}{s_{1x}-\Gamma_{xy}}\tilde{\psi}_{0}=
\tilde{\psi}_{0}+\frac{\Gamma_{xy}}{s_{1x}-\Gamma_{xy}}\tilde{\psi}_{0},
\]
\[
\psi=\tilde{\psi}_{0}+\sum\frac{\Gamma_{xy}}{s_{1x}-\Gamma_{xy}}\left|\psi_{k}\right\rangle \left\langle \psi_{k}|\tilde{\psi}_{0}\right\rangle,
\]
\[
\psi=\tilde{\psi}_{0}+\sum\frac{s_{1xl}}{s_{1x}-s_{1xl}}\left|\psi_{k}\right\rangle \left\langle \psi_{k}\right|\left.\tilde{\psi}_{0}\right\rangle, 
\]
\[
\psi=\tilde{\psi}_{0}+\frac{4\pi}{\epsilon_{2}}\sum\frac{s_{1xl}}{s_{1x}-s_{1xl}}\left|\psi_{k}\right\rangle \left\langle \psi_{k}\right|\tilde{\Gamma}\left.|\rho\right\rangle ,
\]
\[
\psi=\tilde{\psi}_{0}+\frac{4\pi}{\epsilon_{2}}\sum\frac{s_{1xl}}{s_{1x}-s_{1xl}}\left|\psi_{k}\right\rangle \left\langle \psi_{k}\right|\Gamma_{xy}\left.|\rho\right\rangle ,
\]
\[
\psi=\tilde{\psi}_{0}+\frac{4\pi}{\epsilon_{2}}\sum\frac{s_{1xl}^{2}}{s_{1x}-s_{1xl}}\left|\psi_{k}\right\rangle \left\langle \psi_{k}\right|\left.\rho\right\rangle,
\]
\[
\psi=\tilde{\psi}_{0}+\frac{4\pi}{\epsilon_{2}}\sum\frac{s_{1xl}^{2}}{s_{1x}-s_{1xl}}\left|\psi_{k}\right\rangle \nabla\psi_{k}^{*}\left(\boldsymbol{r}_{0}\right)\cdot\boldsymbol{p}.
\]
Now we expand $\tilde{\psi}_{0}$ using the eigenstates of a particle with anisotropy in the $z$ axis:
\[
\tilde{\psi}_{0}=\psi_{0}+\frac{4\pi}{\epsilon_{2}}\sum\frac{s_{1zl}^{2}}{s_{1z}-s_{1zl}}\left|\psi_{k}\right\rangle \nabla\psi_{k}^{*}\left(\boldsymbol{r}_{0}\right)\cdot\boldsymbol{p},
\]
where $\psi_{0}$ is the electric potential for the dipole in free space.
All in all we get:
\[
\psi=\psi_{0}+\frac{4\pi}{\epsilon_{2}}\sum\frac{s_{1zl}^{2}}{s_{1z}-s_{1zl}}\left|\psi_{k}\right\rangle \nabla\psi_{k}^{*}\left(\boldsymbol{r}_{0}\right)\cdot\boldsymbol{p} \]
\[+\frac{4\pi}{\epsilon_{2}}\sum\frac{s_{1xl}^{2}}{s_{1x}-s_{1xl}}\left|\psi_{k}\right\rangle \nabla\psi_{k}^{*}\left(\boldsymbol{r}_{0}\right)\cdot\boldsymbol{p}.
\]

\end{widetext}
\end{document}